\documentclass[traditabstract, twocolumns]{aa}

\usepackage{graphicx}
\usepackage{txfonts}
\usepackage{natbib}
\usepackage{xspace}
\usepackage{xcolor}
\usepackage{scalerel}
\usepackage[normalem]{ulem}
\usepackage{enumitem} 
\usepackage[colorlinks=true,linkcolor=blue,citecolor=blue, urlcolor=blue]{hyperref}
\usepackage{soul}   

\newcommand{\pycs}{{\tt PyCS3}\xspace}

\newcommand{\dynesty}{{\tt DYNESTY}\xspace}
\newcommand{\jax}{{\tt JAX}\xspace}

\newcommand{\Jzeroun}{Q~J0158$-$4325\xspace}

\newcommand{\e}[1]{_{\rm #1}} 
\def \t {\tilde}

\def \l {\left}
\def \r {\right}

\def\kmspc{${\textrm{km}\, \textrm{s}^{-1} \textrm{Mpc}^{-1}}$}
\def\kms{${\textrm{km}\, \textrm{s}^{-1}}$}

\begin{document}

\title{Evidence for a milliparsec-separation supermassive binary black hole with quasar microlensing\thanks{Light curves presented in this paper are only available in electronic form
at the CDS via anonymous ftp to \url{cdsarc.u-strasbg.fr} (130.79.128.5)
or via \url{http://cdsweb.u-strasbg.fr/cgi-bin/qcat?J/A+A/}}, \thanks{Animated Figs. 5 and 9 are available at \url{https://www.aanda.org/10.1051/0004-6361/202244440/olm}.}}

\author{
M.~Millon\inst{\ref{epfl},\ref{stanford}} \and
C.~Dalang\inst{\ref{unige}} \and 
C.~Lemon\inst{\ref{epfl}} \and 
D.~Sluse\inst{\ref{Liege}} \and 
E.~Paic \inst{\ref{epfl}} \and 
J.~H.~H.~Chan \inst{\ref{epfl}} \and
F.~Courbin\inst{\ref{epfl}}  
}

\institute{
Institute of Physics, Laboratory of Astrophysics, Ecole Polytechnique 
F\'ed\'erale de Lausanne (EPFL), Observatoire de Sauverny, 1290 Versoix, 
Switzerland \label{epfl}\goodbreak \and
Kavli Institute for Particle Astrophysics and Cosmology and Department of Physics, Stanford University,
Stanford, CA 94305, USA\label{stanford} \and
Universit\'e de Gen\`eve, D\'epartement de Physique Th\'eorique and Center for Astroparticle Physics, 24 quai Ernest-Ansermet, CH-1211 Gen\`eve 4, Switzerland \label{unige}\goodbreak \and
STAR Institute, Quartier Agora - All\'ee du six Ao\^ut, 19c B-4000 Li\`ege, Belgium \label{Liege} \goodbreak
}

\date{\today}
\abstract{
We report periodic oscillations in the 15-year-long optical light curve of the gravitationally lensed quasar \Jzeroun at $z_s = 1.29$. The signal is enhanced during a high magnification microlensing event of the quasar that the fainter lensed image, B,  underwent between 2003 and 2010. We measure a period of $P\e{o}=172.6 \pm 0.9$ days, which translates to $75.4 \pm 0.4 $ days in the quasar frame. The oscillations have a maximum amplitude of $0.26 \pm 0.02$ mag and decrease concurrently with the smooth microlensing amplitude. We explore four scenarios to explain the origin of the periodicity: (1) the high magnification microlensing event is due to a binary star in the lensing galaxy, (2) \Jzeroun contains a supermassive binary black hole system in its final dynamical stage before merging, (3) the quasar accretion disk contains a bright inhomogeneity in Keplerian motion around the black hole, and (4) the accretion disk is in precession. 
Of these four scenarios, only a supermassive binary black hole can account for both the short observed period and the amplitude of the signal, through the oscillation of the accretion disk towards and away from high-magnification regions of a microlensing caustic. The short measured period implies that the semi-major axis of the orbit is $\sim10^{-3}$ pc and that and the coalescence timescale is $t\e{coal} \sim 1000$ years, assuming that the decay of the orbit is solely powered by the emission of gravitational waves. The probability of observing a system so close to coalescence, in a sample of only 30 monitored lensed quasars, suggests either a much larger population of supermassive binary black holes than predicted or, more likely, that some other mechanism significantly increases the coalescence timescale. Three tests of the binary black hole hypothesis include: (i) the recurrence of oscillations in photometric monitoring during any future microlensing events in either image, (ii) spectroscopic detection of Doppler shifts (up to $\sim$ 0.01$c$) associated with optical emission in the vicinity of the black holes, and (iii) the detection of gravitational waves through pulsar timing array experiments, such as the Square Kilometre Array, which will have the sensitivity to detect the $\sim$100 nano-hertz emission.
}


\titlerunning{Periodicity in quasar microlensing light curves}
\maketitle

\section{Introduction}

The formation of supermassive binary black holes (SMBBHs) is an expected end product that naturally emerges from the hierarchical assembly of multiple galaxy mergers \citep[][]{Haehnelt2002, Volonteri2003}. The binding of the two black holes in the central parsec of the merging galaxies is first driven by dynamical friction until other mechanisms, such as stellar hardening and disk-driven torques, shrink the orbits further \citep[see e.g.][for a review]{LisaCollaboration2022}. Once the SMBBH reaches a separation of the order of 0.01 parsec, the emission of gravitational waves (GWs) efficiently dissipates the angular momentum and the merger of the two black holes becomes inevitable \citep{Begelman1980}.  

The process that leads to the merger of two supermassive black holes (SMBHs) is described in numerical simulations over a wide range of dynamical scales \citep[e.g.][]{Merritt2006, Dotti2007, Cuadra2009} but remains largely unobserved\footnote{To date, OJ 287 is the only confirmed close SMBBH, which was detected from the repeated pairs of outbursts every 12.2 years, interpreted as a secondary black hole crossing the accretion disk of the primary black hole \citep{Valtonen2008}.}.  Measuring the number density of SMBBHs across redshift would improve our understanding of the mechanisms that lead to the formation of black hole pairs, and help refine the expected number of mergers that current and future GW interferometers will detect. The main observational difficulty comes from the insufficient resolution of the imaging surveys, which limits the minimum separation between the detected pairs of active galactic nuclei to a few kiloparsecs \citep[see e.g.][for recent discoveries]{Tang2021, Chen2022, Lemon2022}. The higher resolution of radio observations offers the possibility to detect closer pairs \citep[][]{Rodriguez2006}, but this technique remains limited to the nearest galaxies and to a minimal separation of $\sim10$ pc, leaving the sub-parsec-separation SMBBHs undetected. These systems are, however, the most interesting ones as they are potential sources of GWs in the nano-hertz frequency range. These frequencies fall within the highest sensitivity band of pulsar timing array (PTA) experiments, which means this signal may be observable in the future. Unfortunately, they are also notoriously difficult to detect since their separation is far below the resolution limit of even the largest radio telescopes.

Consequently, candidates have been searched for through indirect techniques, although the observable signature of such close SMBBHs remains an open question \citep{Bogdanovic2008, Shen2010, Montuori2011,Gultekin2012}. Spectroscopic observations can potentially reveal the presence of small-separation SMBBHs through the presence of double-peaked emission lines \citep[e.g.][]{Dotti2009, Bogdanovic2009, Boroson2009} or through a change in the broad line velocities over time \citep{Eracelous2012}, although the displacement of the lines could also be attributed to unusual structures in the broad line region (BLR). With the advent of recent time-domain surveys, candidates have also been proposed from the observed periodicity in some quasar light curves \citep{Graham2015, Liu2016, Charisi2016, Chen2020, ChenY2022, Oneil2022}. With this technique, \cite{Jiang2022} reported a rapidly decaying signal in optical and X-ray light curves, interpreted as the imminent merger of a secondary black hole on a highly eccentric orbit. However, this interpretation is called into question by the recent spectroscopic observations of \cite{Dotti2022}, which rather favour the possibility of a precessing accretion disk to explain the periodicity seen in the optical light curves. This debate illustrates the difficulty of unambiguously identifying the signature of a SMBBH through spectroscopy or spatially unresolved light curves.

In this work we exploit gravitational microlensing to zoom in onto the inner structure of the $z_s = 1.29$ strongly lensed quasar \Jzeroun \citep{morgan99}. This allows us to reveal the presence of a sub-structure in the accretion disk far beyond the resolving power of any other imaging techniques. We interpret this sub-structure as a new candidate SMBBH, with a separation of the order of a milliparsec.

Microlensing is a phenomenon that can occur in strongly lensed quasars when a star from the lens galaxy approaches one of the multiple images of the quasar. In addition to the gravitational lensing effect produced by the entire galaxy, the star itself acts as a gravitational lens, also producing a splitting of the quasar's image. The typical image separation produced by a microlens is of the order of a micro-arcsecond and is thus far too small to be resolved. However, the lensing micro-(de)magnification produced by the star can be detected. As the star passes in front of one of the quasar images, it modulates its magnification, hence producing `extrinsic' variations on top of the `intrinsic' stochastic variations of the quasar. The first detection of extrinsic variability attributed to microlensing is reported in \cite{Irwin1989} in the Einstein Cross (Q 2237$+$0305). This signal is now commonly seen in the light curves of strongly lensed quasars and is a nuisance for time-delay measurements \citep[e.g.][]{Poindexter2007, Tewes2013,  Millon2020b}.

It is a remarkable coincidence that the Einstein radii of the stars acting as microlenses are typically slightly smaller than or are similar to the characteristic angular size of accretion disks \citep{Mosquera2011}. This has an extremely important consequence: as the alignment between the quasar, the star, and the observer slowly changes over time, different regions of the disk are magnified, hence offering the possibility to scan the structure of the accretion disk on nano-arcsecond scales. Microlensing is therefore a unique tool for probing the inner parsec near the central black hole. This method is also highly sensitive to additional structures in the accretion disk, for example mini-disks around a binary companion \citep{Yan2014}.

The COSmological MOnitoring of GRAvItational Lenses (COSMOGRAIL) programme \citep[][]{Courbin2005, Millon2020b} provides the largest dataset to date in which to search for such microlensing events. It consists of a sample of $\sim$ 30 strongly lensed quasar light curves with measured time delays. Once the time delays are measured, the microlensing signal can easily be isolated by shifting the curves by their time delays and subtracting them pair-wise. The resulting difference light curves are therefore free of the intrinsic variability of the quasar and contain only the extrinsic microlensing variations. Most of the COSMOGRAIL systems have been observed for more than 10 years, thus offering a long enough baseline to detect microlensing signatures. Slow microlensing variations (i.e. on a timescale of years) are observed in most of the lensed systems and are often used to set constraints on the accretion disk size \citep[see e.g.][for recent measurements]{Morgan2018, Cornachione2020} or on the temperature profile of the disk \citep{Eigenbrod2008, Goicoechea2020}.

However, several studies have reported that the microlensing signal is in fact much more complex than just a slow modulation of the image magnification \citep[][]{Schild1996,  Hjorth2002, Burud2002, Schechter2003, Millon2020d}. It also contains high-frequency variations (on a timescale of weeks to months) that are too fast to be attributed to stars passing in front of one of the quasar images, unless the stars in the lens galaxy move at relativistic speeds. The fast variations have been tentatively attributed to microlensing by a population of planet-mass microlenses \citep{Schild1996}, variations in the accretion disk size over time \citep{Blackburne2010}, inhomogeneities in the accretion disk \citep{Gould1997, Schechter2003, Dexter2011}, or broad absorption clouds shadowing the quasar \citep{Wyithe2002}. Works by \cite{Sluse2014} and \cite{Paic2021} also propose that a differential magnification of the reverberated flux by the BLR could produce extrinsic variations on the same timescale as the intrinsic variations of the quasar.

In the case of \Jzeroun, the fast microlensing variations appear to be periodic, which is not observed in any other lensed system monitored by COSMOGRAIL. This periodic signal is visible over the period 2003-2010, which coincides with the period where the microlensing magnification of image B is maximal. In this work we aim to qualitatively explain the origin of this periodicity. This paper is organised as follows: In Sect. \ref{sec:observation} we describe the observational data used in this analysis and how the microlensing signal is extracted. Section \ref{sec:model} presents the measurement of the period and amplitude of the periodic signal detected in the difference curve of \Jzeroun with a simple analytical model. Section \ref{sec:explanation} tests different hypotheses regarding the origin of this periodicity. Finally, we conclude with a discussion of our results in Sect. \ref{sec:discussion}. Throughout this paper we convert the angular size into physical size, assuming flat $\Lambda$ cold dark matter (CDM) cosmology with $\Omega_m = 0.3$, $\Omega_{\Lambda} = 0.7$, and $H_0 = 70$ \kmspc. 

\begin{figure}[htpb]
    \centering
    \includegraphics[width=0.48\textwidth]{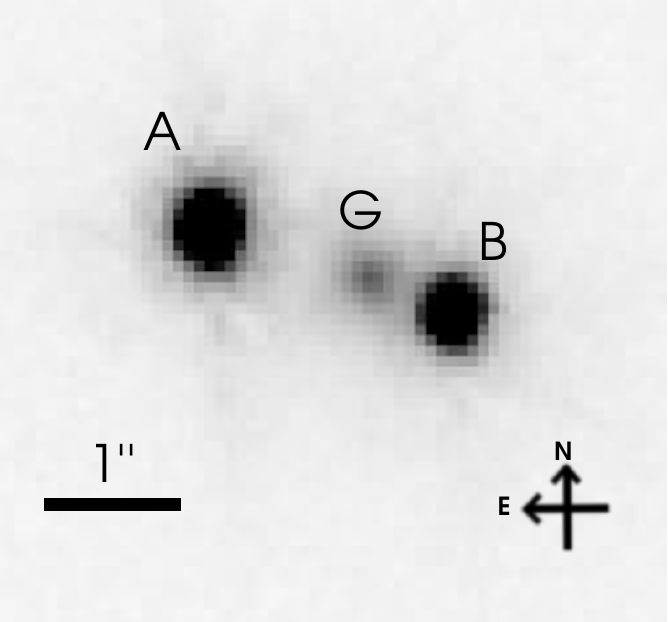}
    \caption{HST image of doubly imaged quasar \Jzeroun in the F814W filter (programme ID 9267; PI: Beckwith).}
    \label{fig:J0158_config}
\end{figure}

\section{Observational data}
\label{sec:observation}
\subsection{Data reduction}

\begin{figure*}[h!]
    \centering
    \includegraphics[width=\textwidth]{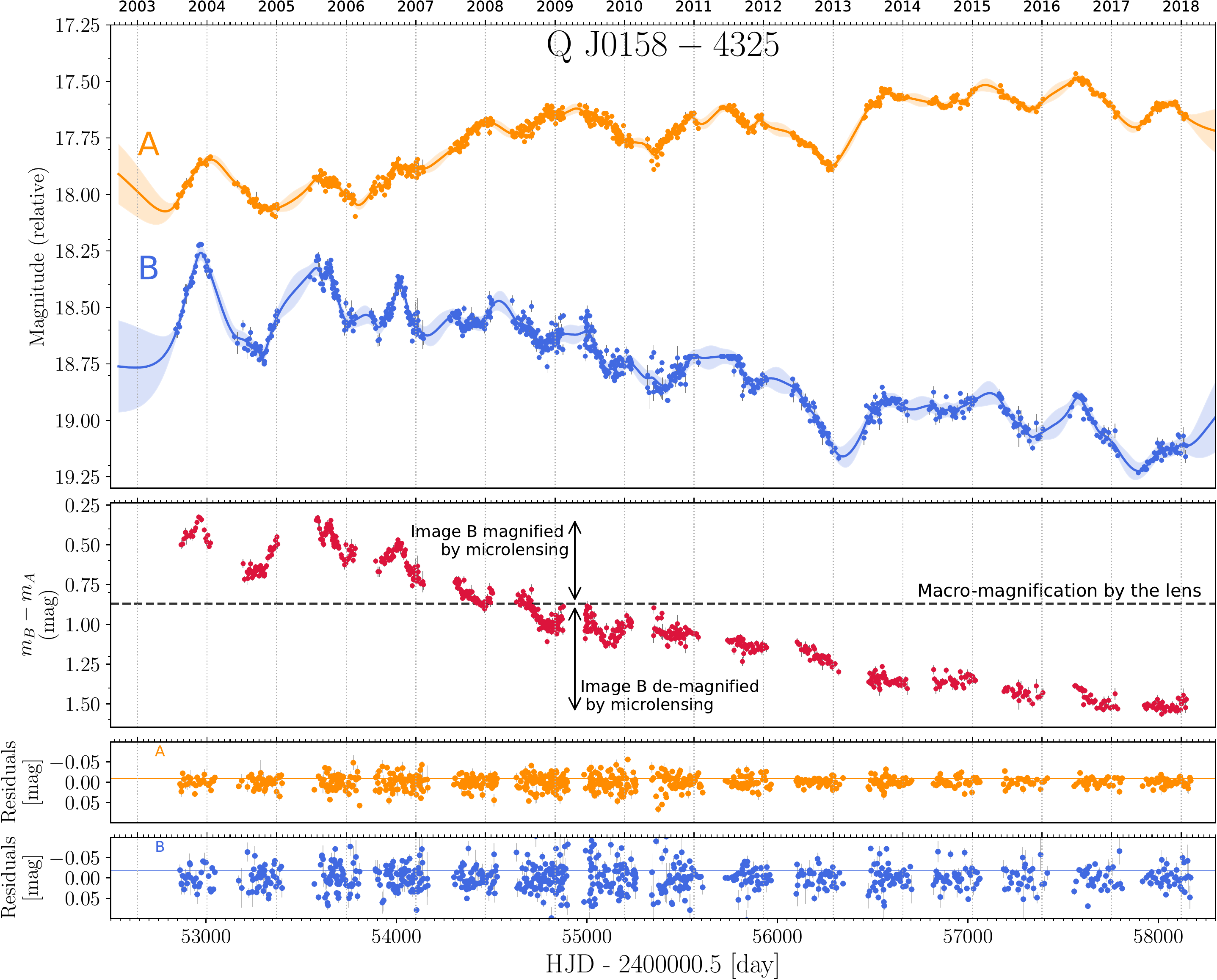}
    \caption{R-band light curve of the lensed quasar \Jzeroun. The light curves combine the data obtained at Euler \citep[2005-2018;][]{Millon2020b} and SMARTS \citep[2003-2010,][]{Morgan2012}. \textit{Top panel:} The solid blue and orange lines correspond to the Gaussian process regression used to interpolate the data, along with their uncertainties (shaded envelope). \textit{Second panel:} Difference light curve between image B and image A, shifted by the time delay. We interpolate between the data points using the Gaussian process regression shown in the top panel. The horizontal dashed line shows the expected flux ratio, in the absence of microlensing, computed from the macro lens model. \textit{Third and fourth panels:} Residuals of the Gaussian process regression.}
    \label{fig:lcs}
\end{figure*}

We use the R-band light curves of the doubly lensed quasar \Jzeroun obtained from 13 years of monitoring at the Leonhard Euler 1.2m Swiss Telescope (hereafter Euler) in La Silla, Chile, in the context of the COSMOGRAIL programme. Figure \ref{fig:J0158_config} shows the lensing configuration of \Jzeroun, as observed by the \textit{Hubble} Space Telescope (HST). The reduction and deconvolution of the Euler images are described in detail in \cite{Millon2020b} and are based on the MCS deconvolution algorithm \citep{Magain1998}. This procedure allows us to precisely extract the flux at the position of the multiple images while removing the contamination from the lens galaxy. The Euler data cover the period August 2005 - February 2018, with 527 epochs. Compared to the data presented in previous publications, the light curves are now calibrated using the star located $\sim2$\arcsec to the east-south-east of the lens, labelled N1 in Fig. A.1 of \cite{Millon2020b}. We used the Dark Energy Survey (DES) Data Release 2 photometry \citep{DES2021} of this star in the R band to compute the zero point of the instrument and calibrate the light curves. We note that this absolute calibration is only approximate due to a possible mismatch between the DES $r$ filter and the RG (`Rouge Genève') filter used for these observations. This does not affect the present work. 

In addition, we complement our dataset with 252 epochs taken between August 2003 and December 2010 at the SMARTS 1.3 m telescope with the ANDICAM optical and infrared camera, published in \cite{Morgan2012}. Since these data overlap with the Euler monitoring campaign, we merge all datasets into a single light curve after fitting a flux and magnitude correction to compensate for the slight photometric offsets mainly due to the differences in the filters and detector responses. This is performed with the curve-shifting package \pycs \footnote{\url{https://gitlab.com/cosmograil/PyCS3}} \citep{Tewes2013, Millon2020c}, which we use to fit a spline model on each image's SMARTS light curve. We then minimise the difference between the spline-interpolated light curves and the Euler data by applying a magnitude shift, followed by a shift in flux. Including both the Euler and SMARTS data, we obtain an interrupted light curve between August 2003 and February 2018 totalling 779 epochs\footnote{Our data are publicly available from the COSMOGRAIL database: \url{https://obswww.unige.ch/~millon/d3cs/COSMOGRAIL_public/}}. 

\subsection{Microlensing curve}

The time delay between image A and image B of \Jzeroun has been measured to be $\Delta t_{AB}=22.7 \pm 3.6$ days from the Euler and SMARTS monitoring data, with image A leading image B \citep{Millon2020b}. The microlensing signal can be extracted by shifting the curves by the estimated time delay and subtracting them. In doing this, we use a Gaussian process regression to interpolate between the data points before performing the subtraction. The resulting difference curve is shown in the second panel of Fig. \ref{fig:lcs}. We refer to this curve as the `microlensing curve' in the rest of this paper, but we note that it contains all extrinsic variations from both images not related to the quasar intrinsic variations. We do not interpolate over season gaps since the Gaussian process regression is poorly constrained in these parts of the light curves. Thus, the seasons of the microlensing curve are 22.7 days shorter than the visibility season. The photometric uncertainties of the microlensing curve are computed by adding in quadrature the photometric uncertainties of image B and the uncertainties of the Gaussian process model fitted onto image A. The uncertainty in the time delay does not significantly impact our microlensing curve as a shift of the time delay by 3.6 days in either direction only introduces an additional error of the order of 3 mmag, that is, $\sim 7$ times smaller than the average photometric uncertainty of the Euler difference light curve. We therefore neglect this additional source of uncertainty. 

The dotted horizontal line on the second panel of Fig. \ref{fig:lcs} indicates the expected magnitude difference from the macro-models of \cite{Morgan2008}, $\Delta_0 = 0.87$ mag. The microlensing curve shows a slow decrease between 2005 and 2018, with image B initially $\sim$ 0.55 mag brighter than predicted by the macro-models published by \cite{Morgan2008}. The brightness of image A increases slightly while that of image B decreases more consistently over the same period. In this particular case, it seems that the microlensing variation is dominant in image B, especially in the first half of the monitoring campaign. This scenario is supported by the spectra of \Jzeroun obtained in 2006 \citep{Faure2009}, which reveal an unusually low contrast between the continuum and the broad lines in image B, which is best interpreted as strong microlensing in that image. For these reasons, we assume in the rest of the paper that the microlensing activity was dominant in image B, whereas image A mostly contains the intrinsic signal of the quasar. By removing the intrinsic variations visible in image A, we assume that we obtain clean observations of the extrinsic microlensing activity happening in image B.  

It is remarkable to observe periodic variations in the first part of the microlensing curve, between 2003 and 2011, also corresponding to a period when image B is highly magnified by microlensing. Over this period, the microlensing magnification varies by about 0.7 mag with modulations of $\sim$ 0.2 mag (peak-to-peak) with a period of $\sim170$ days in the observer frame. The period is significantly smaller than 6 months, which rules out the possibility of a seasonal effect. Moreover, the amplitude of the periodic signal is maximal when the microlensing magnification is also maximal, reaching, for example, $0.26 \pm 0.02$  mag during the season 2005 - 2006. This corresponds to a modulation of the flux of image B by 26\%. The periodicity then disappears after 2011, when image B likely becomes de-magnified. Over the 15 years of our monitoring, the microlensing magnification has changed by 1.22 mag, making \Jzeroun one of the most microlensing-affected systems of the COSMOGRAIL sample. 

\section{Period measurement}
\label{sec:model}

\subsection{Empirical model definition}
We considered a simple model to represent the periodic variations seen in the microlensing curve over the period 2005 - 2010, which become largely attenuated over the period 2010 - 2012. Here, we fit the observed flux ratio between the two images of the quasar, $ F\e{\mu, o}(t)$, rather than the magnitude difference:  

\begin{equation}
    F\e{\mu, o}(t) \equiv \frac{F\e{B}(t)}{F\e{A}(t)}.
\end{equation}

\begin{figure*}[h!]
    \centering
    \includegraphics[width=0.85\textwidth]{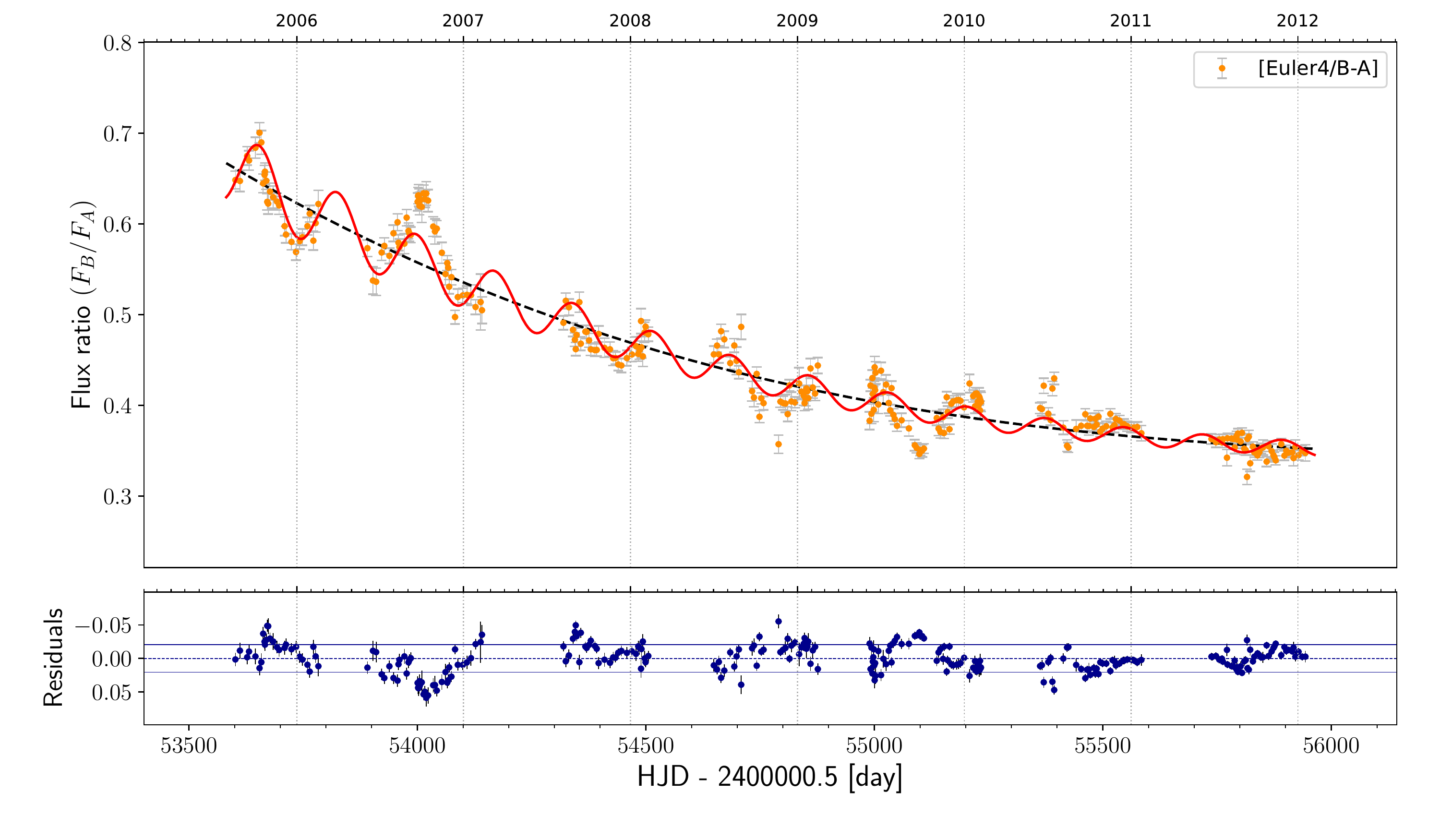}
    \caption{Flux ratio $F_B/F_A$ as observed by the Euler telescope over the period 2005 - 2012. The solid red line shows our best-fit model. The dashed black line shows the smooth polynomial model representing the slow microlensing variations.}
    \label{fig:sinfit}
\end{figure*}

\begin{table*}[htbp] 
\centering
\renewcommand{\arraystretch}{1.5}
\caption{Best-fit reduced $\chi^2$ and median values of the main model parameters for the Euler, SMARTS, and joined Euler+SMARTS datasets. Reported uncertainties correspond to the 16$^{th}$ and 84$^{th}$ percentiles of the posterior distributions. \label{tasb:parameters}}
\begin{tabular}{l|ccccccc} 
 Dataset & $\chi^2\e{red}$ & A & $T$ [days] & $\phi$ & $C$ & $\delta$  & Time span \\ \hline 
SMARTS   & 4.55          &$0.07 ^{+0.02}_{-0.02} $       &$175.83 ^{+1.52}_{-1.54} $        &$2.41 ^{+0.25}_{-0.23} $       &$-0.018 ^{+0.011}_{-0.012} $   &$0.018 ^{+0.001}_{-0.001} $     &2005 - 2012 \\ 
Euler    & 6.46          &$0.11 ^{+0.02}_{-0.02} $       &$172.57 ^{+0.85}_{-0.86} $        &$2.32 ^{+0.15}_{-0.16} $       &$-0.030 ^{+0.009}_{-0.009} $   &$0.019 ^{+0.001}_{-0.001} $     &2005 - 2012 \\ 
Euler+SMARTS     & 6.26          &$0.07 ^{+0.02}_{-0.02} $       &$173.85 ^{+0.90}_{-0.86} $       &$2.49 ^{+0.16}_{-0.15} $       &$-0.018 ^{+0.007}_{-0.008} $        &$0.020 ^{+0.001}_{-0.001} $    &2005 - 2012 \\ 
\end{tabular} 
\end{table*}

We include a smooth model $\mathrm{S}(t)$ for the long-term variation in the microlensing as well as the zero-point flux ratio, described as a third-order polynomial:
\begin{equation}
    S(t) = a_3 t^3 + a_2 t^2 + a_1 t +  a_0,
\end{equation}
which is sufficient to represent the long-term change of the flux ratio over the period 2005 - 2012. Our model considers that the amplitude of the periodic signal is linearly related to the microlensing amplitude:

\begin{equation}
F\e{\mu, m}  = \left(A \cdot \mathrm{S}(t) + C\right) \cdot \sin \left( \frac{2\pi}{P\e{o}}t + \phi \right) + \mathrm{S}(t), 
\end{equation}
where $F\e{\mu, m}$ is the modelled flux ratio, $A$ and $C$ are free scaling parameters, $P\e{o}$ is the period in the observed frame and $\phi$ is the phase.

We perform a Bayesian fit with the likelihood defined as
\begin{equation}
    \ln \mathcal{L} = - \frac{1}{2}\sum_{i=1}^{N} \left( \frac{(F\e{\mu,o}(t_i) - F\e{\mu,m}(t_i))^2}{s_i^2}  +  \ln (s_i^2)\right),
\end{equation}
where $s_i^2 = \sigma_i^2 + \delta^2$, $\sigma_i$ is the individual epoch photometric uncertainty, and $\delta$ is the global intrinsic scatter. We add the intrinsic scatter $\delta$ as a free parameter to account for possibly underestimated photometric uncertainties, or additional complexity in the data not captured by this simple model.

Following Bayes' theorem, we used the posterior distribution of the free parameters, $\vec{\omega}$, 
\begin{equation}
\mathcal{P}(\vec{\omega}| \vec{d}\e{Euler}, \vec{d}\e{SMARTS})  \propto \mathcal{L}(\vec{d}\e{Euler}, \vec{d}\e{SMARTS}|\vec{\omega}) P(\vec{\omega}).
\end{equation}

We chose uninformative flat priors; $A\in[0,5]$, $P\e{o}\in [0,300]$ days, $\phi\in [0,2\pi)$, $C\in[-1,1]$, $\delta \in [0,1]$ and $a_0, a_1, a_2, a_3 \in [-1,1]$. We restrict our analysis to the period 2005 - 2012 where the periodic variations are the most prominent and clearly seen above the noise level. We leave the interpretation of the complex microlensing signal over the period 2003-2005 for the discussion in Sect. \ref{sec:discussion}.

\subsection{Results}
\label{subsec:resutls}

The posterior distributions are sampled using the nested sampling python package \dynesty \citep{Speagle2020}. The median as well as the 16$^{th}$ and $84^{th}$ percentiles of the marginalised posterior distributions are quoted in Table \ref{tasb:parameters} for the SMARTS, the Euler and the joined SMARTS-Euler dataset. Here, the reported $\chi\e{red}^2$ only include the photometric uncertainties. The derived periods from the three datasets are compatible within 2$\sigma$. Our most precise estimation is from the Euler dataset with $P\e{o}=172.6^{+0.9}_{-0.9}$ days. The best-fit to the SMARTS data has a significantly smaller $\chi\e{red}^2$ but mostly because of larger photometric uncertainties. This results in a degraded precision on the derived period when adding this dataset. For this reason, we restrict our analysis to the Euler data only for the rest of this paper. The best-fit on the Euler dataset is shown in Fig. \ref{fig:sinfit}. 

The $\chi\e{red}^2$ of the fit is significantly above 1 for all three datasets, indicating that our single sinusoid, whose amplitude is linearly related to the microlensing magnification, is not sufficient to capture the full complexity of the signal. This is also reflected in the intrinsic scatter, which is significantly larger than 0. We experimented with higher-order corrections of the amplitude of the sinusoid without obtaining a significantly better fit. We thus decided to keep our model as simple as possible but this might be an indication that more complex phenomena, such as the differential reverberation proposed by \cite{Paic2021}, are happening.

\subsection{Lomb-Scargle periodogram}

\begin{figure}[htbp]
    \centering
    \includegraphics[width=0.5\textwidth]{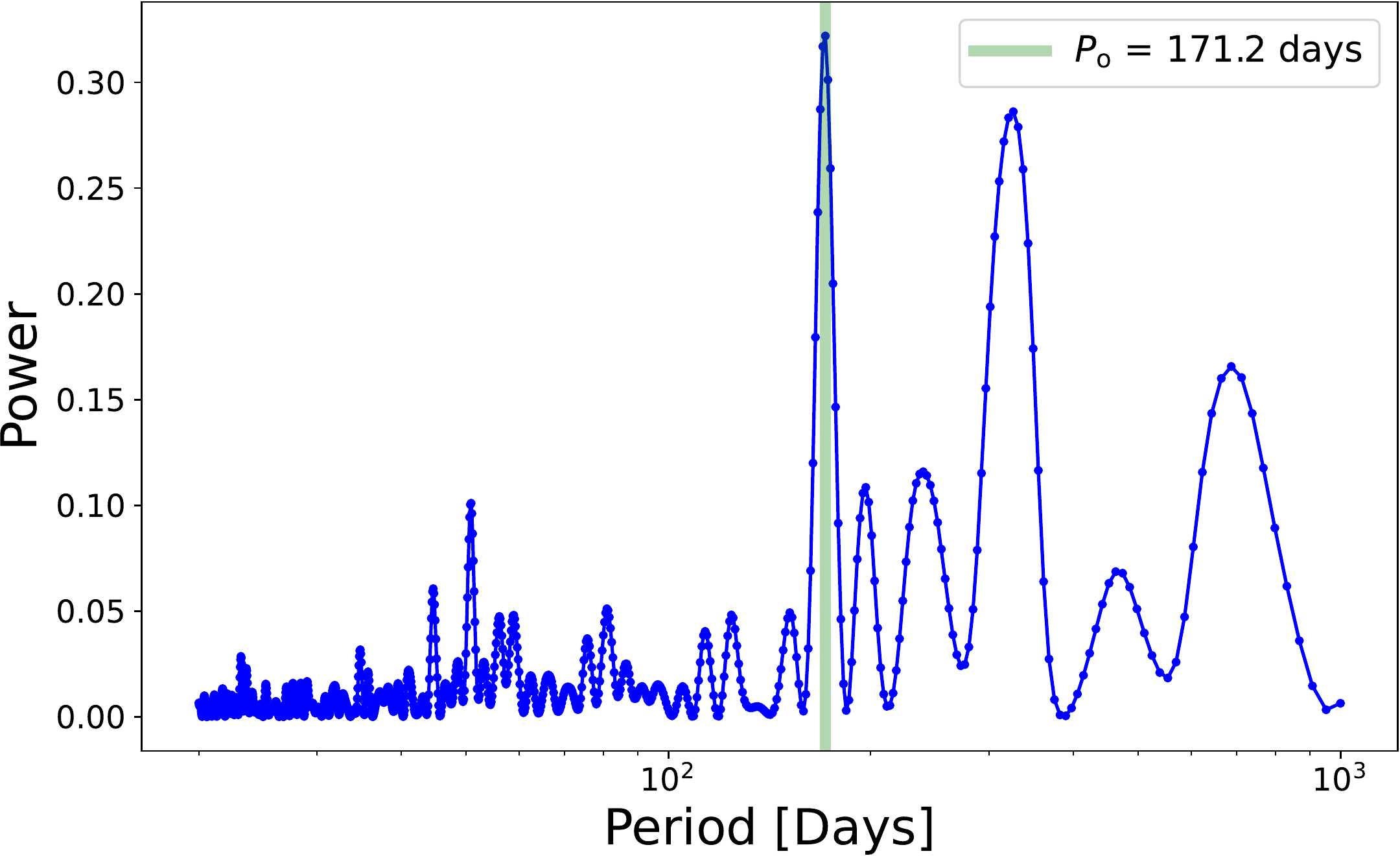}
    \caption{Generalised Lomb-Scargle periodogram of the flux ratio observed by the Euler telescope over the period 2005 - 2011. The data were corrected from the long-term microlensing trend before computing the periodogram. The vertical green line indicates the peak frequency.}
    \label{fig:Lomb-Scargle}
\end{figure}

A standard technique for the spectral analysis of unevenly spaced time series is the Lomb-Scargle periodogram \citep{Lomb1976, Scargle1982}. Here, we used the {\tt PyAstronomy}\footnote{\url{https://github.com/sczesla/PyAstronomy}} \citep{pyastronomy} implementation of the generalised Lomb-Scargle (GLS) periodogram \citep[see][]{Zechmeister2009}, which accounts for offsets and variable uncertainties across the data points. We first corrected our data from the best-fit polynomial found in the previous section in order to remove the long-term microlensing trend. We then applied the GLS algorithm to the corrected data, with frequencies ranging from $20^{-1}$ to $1000^{-1}$\ days$^{-1}$. The resulting periodogram on the Euler dataset is displayed in Fig. \ref{fig:Lomb-Scargle}, which shows a clear peak at a period of about 171 days as well as smaller harmonic peaks at about 342 and 684 days. The same three peaks are also clearly detected on the SMARTS and the joined Euler-SMARTS datasets.

Red-noise like variability observed in active galactic nuclei has been shown to potentially produce spurious periodic signals, if only a few cycles are observed \citep{Vaughan2016}. In the present case, we stress that we observe more than ten cycles between 2005 and 2011. Moreover, the periodic signal is not obvious in the direct emission but is unveiled only in the microlensing light curve, where the intrinsic variations of the quasars are expected to be cancelled. We therefore do not expect the microlensing light curve to be affected by the red noise variability of the quasar. However, \cite{Sluse2014} have proposed a mechanism where the stochastic variability of the quasar could be echoed in the microlensing curve if a significant fraction of the R-band flux is emitted from a region unaffected by microlensing. We show in Appendix \ref{Appendix:A} that this mechanism is not sufficient to reproduce the large amplitude of the extrinsic variations seen in the microlensing light curve. We still used this physically motivated model to generate 5000 simulated light curves from a damped random walk (DRW) and compute the microlensing curve for each of them, using the same differential microlensing model as presented in \cite{Paic2021}. The simulated light curves have the same sampling and photometric noise as the real data (see Appendix \ref{Appendix:A} for the details of this test).

Only 0.6\% of the curves produces a peak in the GLS periodogram with more power than observed in the Euler data over the period 2005-2011. We conclude that this differential reverberation model is unable to reproduce the periodicity observed in the first part of our observations at 3.7$\sigma$ significance level. Although it might still explain the small amplitude flickering seen in the second part of our observations and in other systems of the COSMOGRAIL sample, we conclude that it is improbable that the observed periodicity arises by chance from the differential reverberation model proposed by \cite{Sluse2014} and \cite{Paic2021}. 

\section{Origin of the periodic signal}
\label{sec:explanation}
We propose four hypotheses to explain the periodicity observed in the extrinsic variability of image B. 

\textbf{Hypothesis 1:} The microlensing magnification is modulated by a secondary star (or a planet) in the lens plane. The microlensing event seen in image B is in fact produced by a pair of microlenses. 
    
    \textbf{Hypothesis 2:} \Jzeroun is a binary black hole, with two SMBHs in their final stage before merging. The motion of the accretion disk around the centre of mass of the system in the source plane is at the origin of the observed signal. 
    
    \textbf{Hypothesis 3:} The accretion disk contains an inhomogeneity in Keplerian motion around the central SMBH, which is approaching the micro-caustic periodically.
    
    \textbf{Hypothesis 4:} The inner part of the accretion disk is in precession. This precession could be due to the Bardeen-Peterson effect \citep{Bardeen1975} or because the disk is eccentric, which implies that the pericentre of elliptical orbits advances at each revolution in a Schwarzschild potential.

Each of these scenarios is detailed in the following subsections, where we propose simple toy models to evaluate if these hypotheses could reproduce the same amplitude and period of the microlensing signal. We assume that the light intensity profile of the quasar's accretion disk is represented by a non-relativistic thin-disk profile \citep{Shakura1973} such that 

\begin{equation}
    \label{eq:thin-disk}
    I_0(R) \propto \frac{1}{\exp(\xi(R)) - 1 },
\end{equation}
where 

\begin{equation}
    \xi(R) = \left( \frac{R}{R_{0}} \right)^{3/4} \left( 1 - \sqrt{\frac{R\e{in}}{R}} \right)^{-1/4}.
\end{equation}

In this last equation, $R\e{in}$ corresponds to the radius of inner edge of the accretion disk and $R_{0}$ is the scale radius, which can be estimated from the black hole mass, $M\e{BH}$: 
\begin{equation}
   R_{0} = 9.7 \times 10^{15} \mathrm{cm} \left( \frac{\lambda_s}{\mu \mathrm{m}} \right) ^{4/3} \left(\frac{M\e{BH}}{10^9 M_{\odot}}\right) ^{2/3} \left( \frac{L}{\eta L\e{E}}\right) ^{1/3},
\end{equation} 
where $\lambda_s$ is the observed wavelength in the quasar rest frame, $L\e{E}$ is the Eddington luminosity and $\eta$ is the accretion efficiency. Assuming a typical Eddington ratio $L/L\e{E} = 1/3$, accretion efficiency of 10\% ($\eta$=0.1), and a black hole mass of $M\e{BH}=1.6 \times 10^{8} M_{\odot}$ based on Mg II line width measurement \citep[$\sim$ 0.35 dex uncertainties,][]{Peng2006a}, we derive a characteristic scale of the accretion disk of $R_0 = 7.9 \times 10^{14}$ cm at 650 nm in the observer frame, corresponding to $\lambda_{s} = 650\mathrm{nm} / (1 + z_s) = 284$ nm in the quasar rest frame. This corresponds to 0.3 light-days. $R_0$ is related to the half light radius of the profile through the simple relation $R_{1/2} = 2.44R_0$. Finally, we fixed
\begin{equation}
    R\e{in} = 6 r_g
,\end{equation}
where $r_g = GM\e{BH}/c^2$ is the gravitational radius. $R\e{in}$ corresponds the size of the innermost stable circular orbit (ISCO) for a Schwarzschild black hole. 
We adopted a fiducial macro lens model from \cite{Morgan2008} for a stellar mass fraction $f\e{M/L} = 0.9$ ($\kappa = 0.72$, $\gamma = 1.03$, $\kappa_{\star} / \kappa = 0.92$ for image B). For the population of    microlenses used in Sect. \ref{subsec:hotspot}, we made similar assumptions to those by \cite{Paic2021}, that is, a Salpeter initial mass function with mean stellar mass $\langle M \rangle = 0.3M_{\odot}$ and a mass ratio of 100 between the heaviest and the lightest microlenses. The mean Einstein radius $R\e{E}$ of the microlenses, projected into the source plane, is defined as\begin{equation}
    R\e{E} = D_s \times \sqrt{\frac{4G\langle M\rangle}{c^2} \frac{D_{ls}}{D_{l}D_s}},
\end{equation}
where $D_s$, $D_{l}$, and $D_{ls}$ are the angular diameter distances to the source, to the lens and between the lens and the source. For $\langle M \rangle = 0.3M_{\odot}$, the Einstein radius is $R\e{E} = 3.4\times 10^{16}$ cm (13.1 light-days). We note that the size of the accretion disk is smaller than the typical Einstein radius of the microlenses ($R_0 / R\e{E} =  0.023$), which makes this system likely to be affected by large microlensing variations. 


\begin{figure*}[h!]
    \centering
    \includegraphics[width=0.9\textwidth]{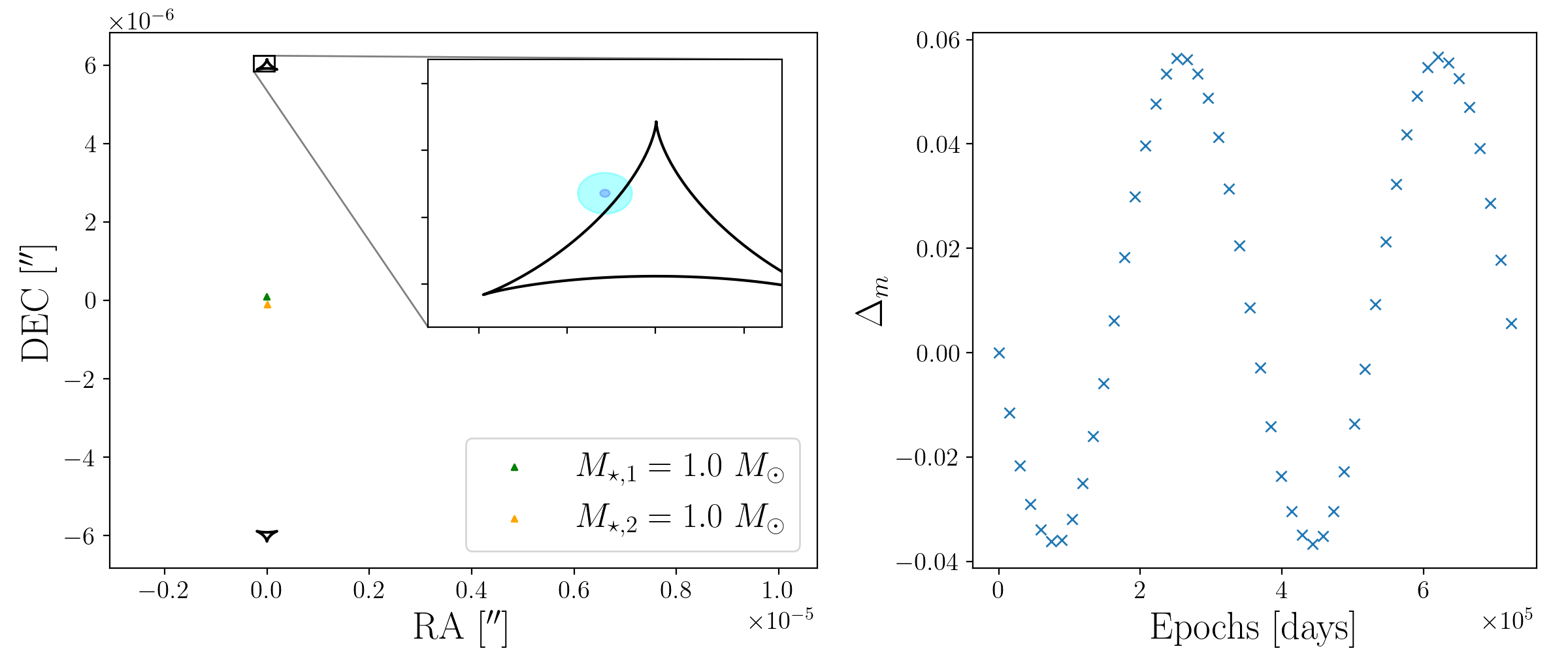}
    \caption{Simulation of the microlensing effect produced by a binary star. \textit{Left panel:} Source plane micro-caustics (black) created by a pair of stars located at the position of image B in the lens plane. The green and orange triangles show the stars' locations, which are separated by 200 AU. The inset panel zooms in onto the position of the accretion disk. The light (dark) blue circle corresponds to the accretion disk size $R_0$ (size of the ISCO, $R_{\mathrm{ISCO}}$). 
    \textit{Right panel:} Magnitude change of image B due to the periodic motion of the microlenses. An animated version of this figure is available at this \href{	https://www.aanda.org/articles/aa/olm/2022/12/aa44440-22/aa44440-22.html}{link}.
    }
    \label{fig:source}
\end{figure*}

\subsection{Binary microlenses}
\label{subsec:planet}

In this first scenario, we assume that the periodic variations originate from a stellar binary (or a planetary system) in the lens plane. We aim at estimating the amplitude of the microlensing modulations that such a binary system would produce. 
First, we fixed the orbital period of the binary system in the lens plane to 

\begin{equation}
    \label{eq:period}
    P_l = 2P\e{o} / (1 + z_l),
\end{equation}where $P\e{o}$ is the measured period in the observer frame and $z_l=0.317$ is the lens redshift. For the measured $P\e{o} = 172.6$ days, this gives $P_l = 262.1$ days. The factor of 2 introduced in Eq.\,\eqref{eq:period} comes from the fact that a binary system produces a modulation of the microlensing signal at half the orbital period. By fixing the period and the masses of the two binary stars, the semi-major axis is imposed through Kepler's second law. Additionally, we assume that the orbital motion is circular and perpendicular to the plane of the sky.  

Second, we use the lens modelling software {\tt lenstronomy} \citep{lenstronomy2021} to generate the microlensed images of the accretion disk. Our lens model is composed of two point masses (representing the stars) plus external convergence and shear ($\kappa =  0.72$, $\gamma = 1.03$) corresponding to the value of our fiducial macro lens model at the position of image B. We note that, since image B is a saddle point, the caustic produced by the pair of stars is split in two, as can be seen on the left panel of Fig. \ref{fig:source} \citep[see e.g.][for a discussion of the properties of microlensing caustics near a macro saddle point]{Schechter2002}. 

We assume a thin-disk profile (Eq.\,\eqref{eq:thin-disk}) located at a distance $d = 0.5 R_0$ from the fold of the caustics in the source plane. We let the system evolve for one full period and compute the total flux of image B at each time step. We compute $\Delta_m$, the maximum peak-to-peak amplitude (in magnitude) of the periodic microlensing signal. It corresponds to the maximal change of microlensing magnification due to the orbital motion of the two stars acting as microlenses. 

The choice of $d = 0.5 R_0$ maximises the amplitude of the periodic signal for a pair of 1$M_{\odot}$ stars. This optimal distance slightly varies with the mass of the microlenses but we fix it to $d = 0.5 R_0$ for all microlenses' masses since it does not change $\Delta_m$ by more than a factor of 10. Similarly, we chose a location near the caustic's fold that maximises the signal but other choices (e.g. positioning the disk near the caustic's cusp) reduces the amplitude by no more than a factor of 10.

\begin{figure*}[h!]
    \centering
    \includegraphics[width=0.49\textwidth]{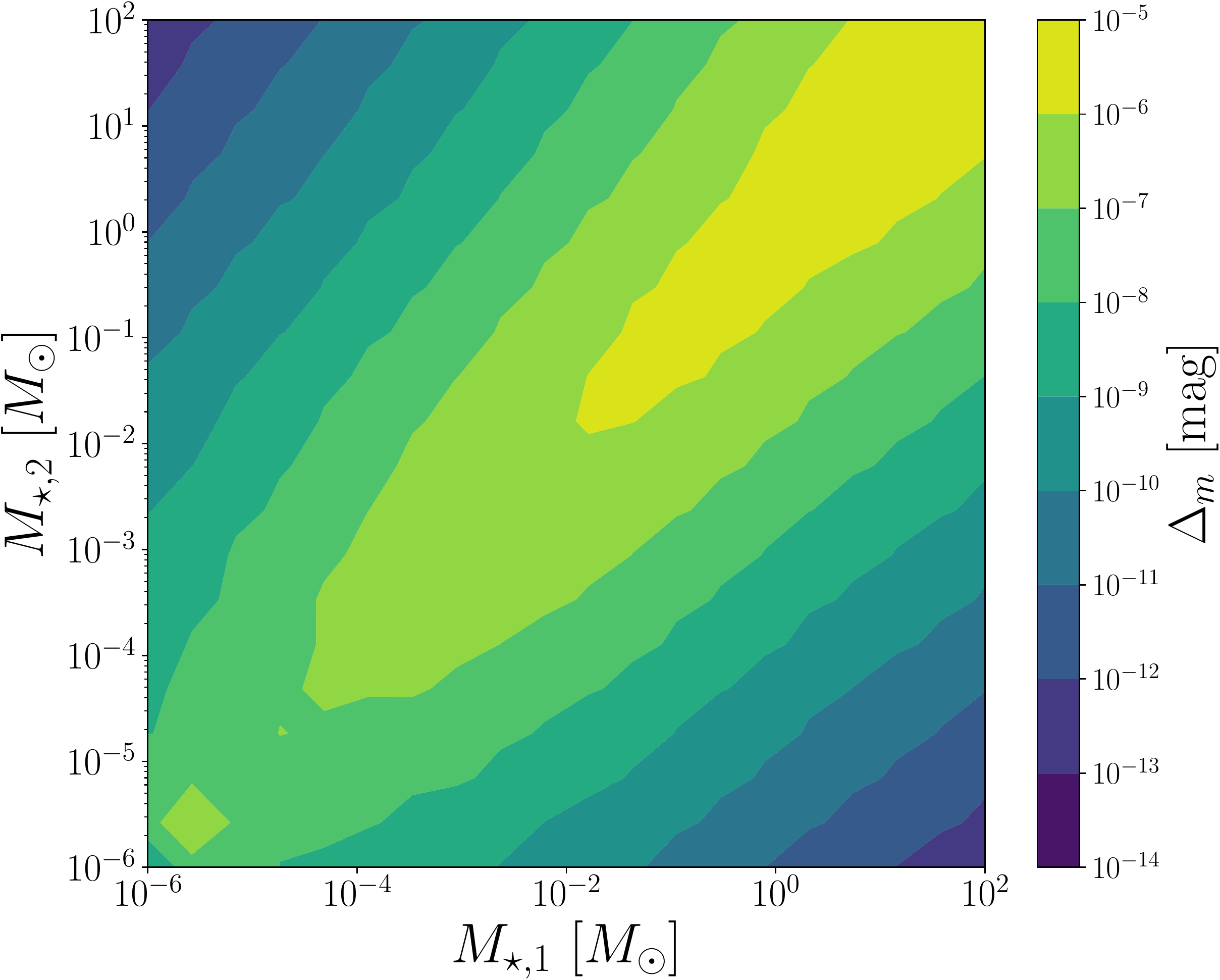}
    \includegraphics[width=0.49\textwidth]{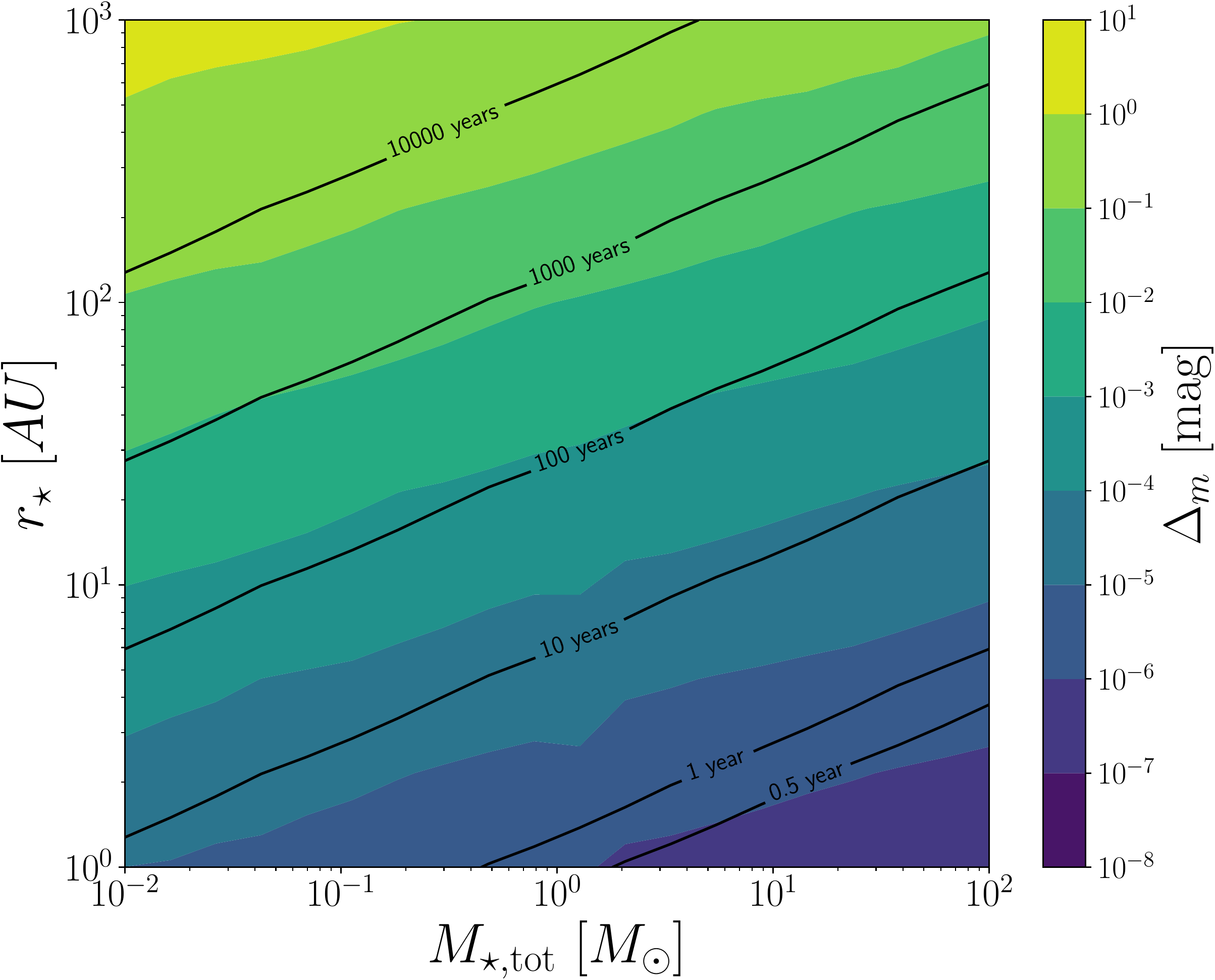}
    \caption{Maximal amplitude of the periodic signal, $\Delta_m$, for a pair of stars of mass $M_{\star,1}$ and $M_{\star,2}$ (left panel). The period is fixed to the observed $P_l = 262$ days, which imposes the separation between the two stars.
    The right panel shows the maximal amplitude of the periodic signal, $\Delta_m$, as a function of the total mass of the binary system, $M_{\star, tot}$, and the radius of the orbit, $r_{\star}$. In this case, the orbital period is not fixed and is indicated by the black contours.}
    \label{fig:planet}
\end{figure*}

Keeping the same source position relative to the caustic, we repeat the experiment for a pair of compact bodies (stars or planets) with a mass $M_{\star,1}$ and $M_{\star,2}$ in the range $10^{-6} - 10^2 M_{\odot}$. The amplitude of the microlensing signal $\Delta_m$ produced by such a binary system is shown on the left panel of Fig. \ref{fig:planet}. We recall that the observed $\Delta_m$ for \Jzeroun is $\sim$ 0.2 mag. Even an extremely massive pair of 100$M_{\odot}$ stars would not be able to produce a periodic modulation of more than $10^{-5}$ mag, that is, 4 orders of magnitude smaller than the observed signal. 

The right panel of Fig. \ref{fig:planet} shows $\Delta_m$ as a function of the orbiting radius $r_{\star}$ of the binary system and its total mass $M_{\star, tot} = M_{\star, 1} + M_{\star, 2}$, with $M_{\star, 1}=M_{\star, 2}$. In this case, the orbital period is not forced to match the observed one. 

In this last example, we are able to reproduce the observed amplitude but not the correct period. This is highlighted in Fig. \ref{fig:source} for a pair of 1 $M_{\odot}$ stars. The modulation of the microlensing amplitude can reach 0.1 mag but only when the two stars are separated by 200 astronomical units (AU). This corresponds to a much longer period of 2000 years. It is therefore not possible to reproduce both the period and amplitude of the observed signal. Even when choosing an ideal source position relative to the micro-caustic and extremely massive microlenses the observed signal is at least 4 orders of magnitude larger than our simulations. Moreover, we took a conservatively small value for the accretion disk size, which might in fact be several times larger than predicted by the thin-disk theory \citep[see e.g.][for an overview of this issue]{Pooley2007, Morgan2010, Cornachione2020b}. A larger disk would further reduce the amplitude of the microlensing signal. It is therefore extremely unlikely that the periodicity observed in the microlensing curve of \Jzeroun originates from binary microlenses in the lens plane. 

\subsection{Supermassive binary black hole}
\label{subsec:BBH}

We explore the possibility that the modulation of the microlensing signal originates from a system composed of two gravitationally bound SMBHs. The orbital period in the source plane is $P_s = P\e{o} / (1 + z_s) = 75.4$ days. We repeat the experiment presented in Sect. \ref{subsec:planet}, but with a single 1$M_{\odot}$ star acting as a microlens in the lens plane. This time, the periodic motion is generated by displacing the centre of the thin-disk profile in the source plane around the centre of mass of the binary system. 

\begin{figure}[h!]
    \centering
    \includegraphics[width=0.5\textwidth]{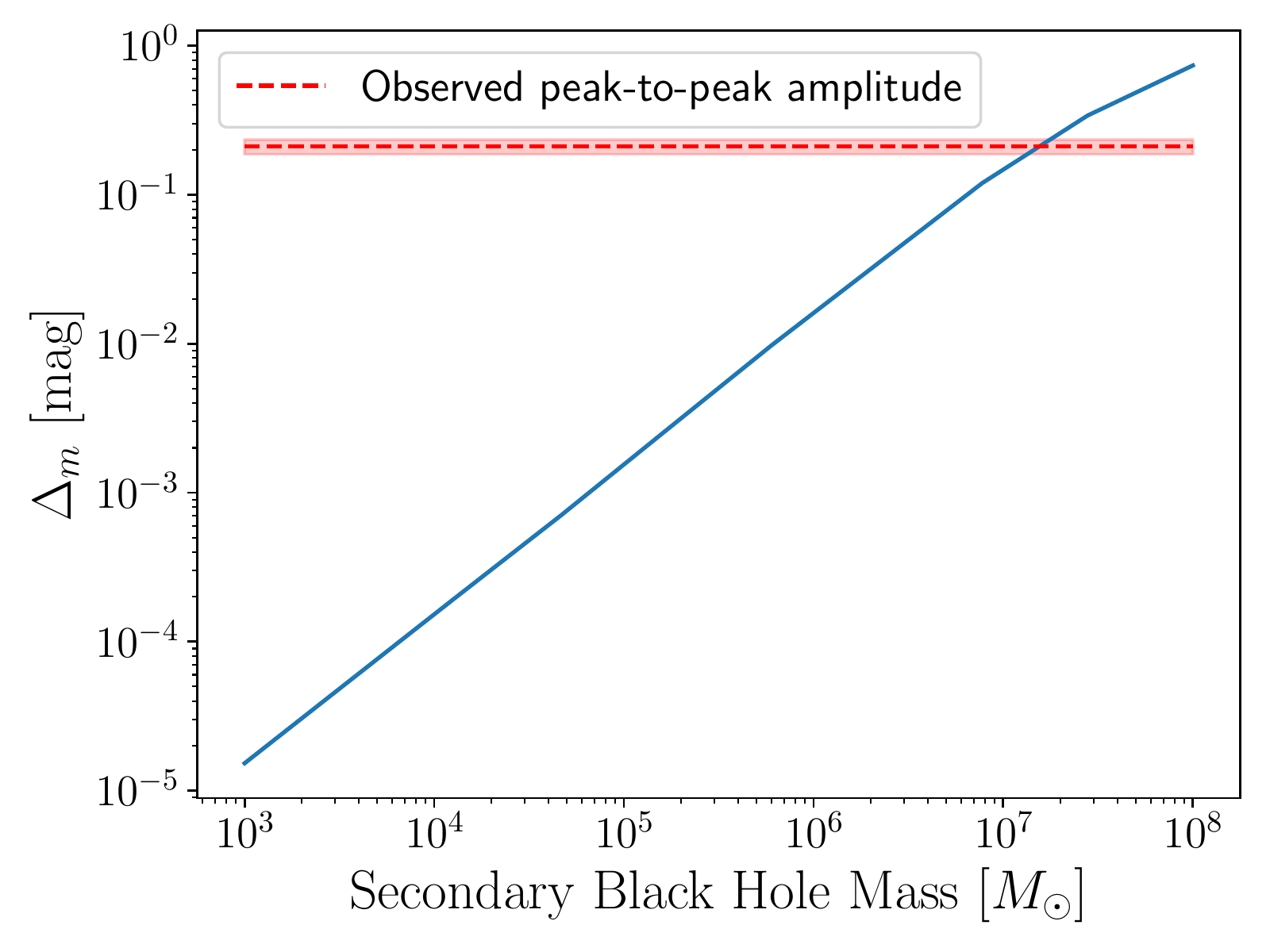}
    \caption{Maximum amplitude of the periodic signal, $\Delta_m$,  as a function of the secondary black hole mass, $M\e{2}$. The primary black hole mass is fixed to our fiducial black hole mass estimate, $M\e{BH} = 1.6 \times 10^8 M_{\odot}$. }
    \label{fig:BBH}
\end{figure}

Similarly to Sect. \ref{subsec:planet}, we position the centre of mass of the system at a distance $d = 0.5 R_0$ from the fold of the micro-caustic to maximise the microlensing amplitude. However, we do not associate any light emission with the secondary black hole; the modulation of the light profile occurs only because the primary black hole and its accretion disk orbit around the system's centre of mass. Here, we neglect the possibility that the secondary black hole may also have its own accretion disk, or that complex structures such as circumbinary disks and mini-disks may arise from the gravitational interaction of the two systems (see also Sects. \ref{subsec:hotspot} and  \ref{sec:discussion} for a more detailed discussion of this issue). Several numerical simulations \citep[e.g.][]{Cuadra2009, Dorazio2013, Bowen2018} predict the formation of a gap between the circumbinary disk and the two spiralling black holes in the centre but the implementation of these profiles is left for future work. Our simple representation is in fact sufficient to reproduce the main features of the microlensing curve, namely the period and the amplitude. Figure \ref{fig:BBH} shows the maximal peak-to-peak amplitude $\Delta_m$ for a secondary black hole's mass in the range $10^3$ - $10^8$ $M_{\odot}$ with the orbital period kept fixed to $P_s$, and the mass of the main black hole $M_1$ fixed to the fiducial black hole mass of $1.6 \times 10^8 M_{\odot}$. 

The observed microlensing amplitude is reproduced for a binary companion mass of $M_2 \sim 10^7 M_{\odot}$. This value should rather be considered as a lower limit for $M_2$ than a proper measurement because we chose an optimal location of the system relative to the micro-caustic and a conservatively small accretion disk size. By changing some of the arbitrarily fixed parameters in the model (such as the mass of the star acting as a microlens or the distance from the caustic), one can easily accommodate bigger masses for the secondary black hole. 

Solely based on our simulations, the hypothesis of a binary SMBH is plausible and reproduces well the observed microlensing curve. If we consider that the accretion disk might move away from the caustic due to the relative motion of the source, the star in the lens galaxy, the lens galaxy itself and the observer, the microlensing magnification would be gradually reduced, and the damping of the signal is easily reproduced. 

However, such binary systems are thought to be rare because of the rapid decay of their orbit due to GW emission. Neglecting the dynamical friction and considering that the system loses energy only through GW emission, the two black holes will eventually merge on a coalescence timescale that depends on the initial eccentricity and semi-major axis of the orbit. Assuming $M_1=1.6 \times 10^8 M_{\odot}$, $M_2 = M_1/10$, and a circular orbit, we obtain an orbiting radius of $r = 9.8\times 10^{-4}$ pc from the observed period of 75.4 days in the source frame. The coalescence time of such a close binary system reads \citep[][]{Peters1964}
\begin{equation}
\label{eq:t_coal}
t\e{coal} = \frac{5}{256} \frac{a^4 c^5}{G^3M_1M_2(M_1+M_2)}\,
\end{equation}
and is thus estimated to be $t\e{coal} \sim  10^3$ years. This timescale is extremely small compared to the age of the quasar at redshift $z_s = 1.29$, which is about 4 Gyr if the quasar was formed around redshift 7. The probability of observing this system in the last $\sim$ 1000 years before merging is approximately $2.6 \cdot 10^{-7}$. If we consider that the black hole could have encountered several merger events during its lifetime, this probability can be increased by a factor of a few but remains very small. Merger rates of SMBBHs are yet to be constrained by observations and depend on the existence of primordial black holes \cite[][see also \citet{Erickcek:2006xc} for an estimation of SMBH merger rates observable with the Laser Interferometer Space Antenna]{LISACosmologyWorkingGroup:2022jok}. 

Nevertheless, it is quite surprising, given the $\sim$30 lensed quasars of the COSMOGRAIL sample, that we observe such a system. However, if the mass of the black hole turns out to be overestimated by a factor of 10, which is possible given that black hole mass estimates based on broad line-width measurements are notoriously uncertain, the coalescence time would be much larger. Uncertainties on the black hole mass are typically of the order of 0.3-0.4 dex from the intrinsic scatter of the virial relationships \citep[e.g.][]{Peterson2004, Restrepo2016} but several biases may affect the measurements, especially if the black hole is a binary. If the mass of the primary black hole is rather of the order of $M_1=1.6 \times 10^7 M_{\odot}$, the same mass ratio would also reproduce both the amplitude and period of the signal. In this case, we find a larger coalescence time of $5 \times 10^4$ years, under the same assumptions. The probability of observing this system would still be small (of the order of $10^{-5}$) and reaches $3 \times 10^{-4}$, if we consider the 30 light curves of the COSMOGRAIL sample.

In summary, the scenario of a SMBBH is appealing to explain the observed signal in the light curve but it hardly accommodates for the very short lifetime of these systems when the decay of the orbit is dominated by GW emission. The probability of observing this system in its final stage before merging is small unless the black hole mass of \Jzeroun is largely overestimated, or if the merging is delayed by the gravitational interaction of a gaseous circumbinary disk (this issue is discussed in more detail in Sect. \ref{sec:discussion}). In this case, detecting such a signal would be rare but not completely excluded. 

Assuming that the emission closest to the larger black hole is not disrupted by the merger process, a possible observational signature would be periodic Doppler shifts of the electromagnetic emission. This would be observed at X-ray wavelengths, by measuring the 6.4keV Fe-K$\alpha$ line shift. The Keplerian velocity of the system is $\sim$ 27,000 \kms, but for a mass ratio of $q\sim$ 10, the line-of-sight velocity of the primary black hole is up to 2700 \kms, depending on the inclination angle. Line energy variations, both intrinsic and extrinsic \citep{Bhatiani2019}, are typically larger than the 5\% level, making such a periodic spectroscopic detection difficult. For this reason, we cannot convincingly conclude that the Fe-K$\alpha$ line shift seen in the X-ray monitoring data obtained for this system by \cite{Chartas2017} is due to a secondary black hole. Continued photometric monitoring at optical wavelengths, however, should clearly reveal the periodic signal during a microlensing event in either image, under the SMBBH hypothesis.

\begin{figure*}[h!]
    \centering
    \includegraphics[width=\textwidth]{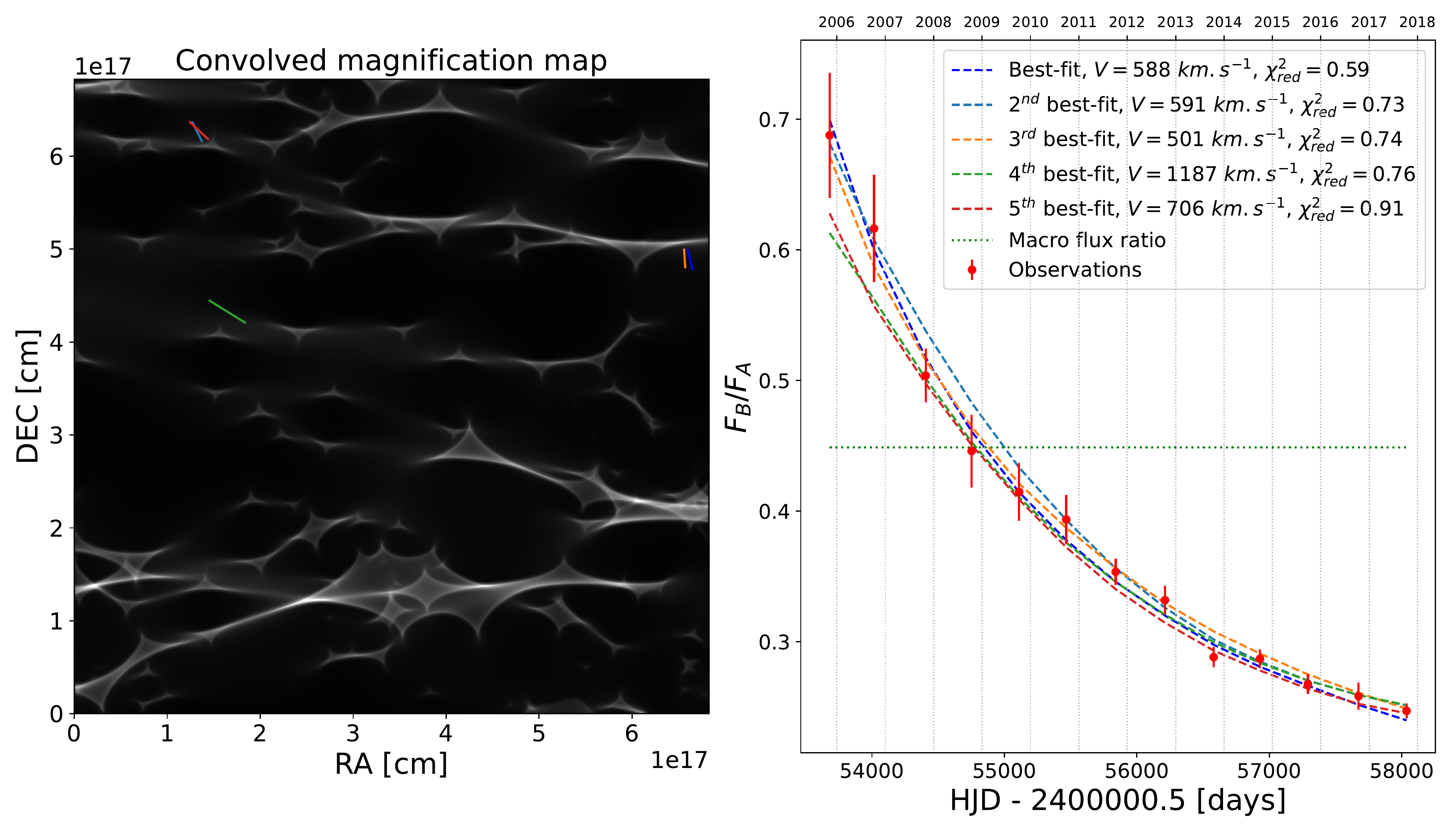}
    \caption{Modelling of the long-term flux ratio variations from realistic microlensing simulations. \textit{Left panel:} Magnification map, convolved by a thin-disk light profile with $R_0 = 7.9 \times 10^{14}$ cm (0.3 light-days). The five best-fit trajectories are shown in colours. \textit{Right panel:} Observed flux ratios $F_B / F_A $ of the two lensed images. The observations are averaged by season in order to fit only the long-term variations, attributed to the displacement of the quasar through the micro-caustic network. The five best-fit trajectories are shown as dashed lines. The horizontal dotted line corresponds to the flux ratio expected from the macro lens model. The legend indicates the total transverse velocity, $V$, corresponding to the selected trajectory as well as the associated reduced $\chi^2$ of the fit. }
    \label{fig:trend_fit}
\end{figure*}
\subsection{Inhomogeneities in the accretion disk around the ISCO}
\label{subsec:hotspot}

In our third scenario, we explore the possibility of a small inhomogeneity in the accretion disk, differentially amplified due to microlensing magnification of image B. In this case, the bright `hotspot' may periodically approach a micro-caustic, hence modulating the magnification of this region of the disk. To test this scenario, we generate microlensing magnification maps by inverse ray-shooting with the GPU-D software \citep{Vernardos2014}. The size of the maps is $20 R\e{E} \times 20R\e{E}$.

First, we searched for trajectories in the magnification maps that correspond to the long-term trend observed in the microlensing curve. This is performed in a similar way to the Monte-Carlo method presented in \cite{Kochanek2004}, except that we are not aiming to measure the accretion disk size of the quasars, which is degenerate with the total relative velocity between the quasar, the microlenses and the observer. Here, we fix the accretion disk size to its thin-disk predicted value, $R_0 = 7.9 \times 10^{14}$ cm (0.3 light-days). We generate $10^6$ random trajectories through the magnification maps assuming a total transverse velocity $V$, in the range [0-2000] \kms. We compute the flux ratios between image A and image B along these trajectories, assuming that A is unaffected by microlensing. We compare these simulations to observational data by taking the weighted mean of the observed flux ratios in each season. The five best-fit trajectories are shown in Fig. \ref{fig:trend_fit}.

Several combinations of total transverse velocity and location relative to the micro-caustics offer a good fit to the global microlensing trend with a $\chi\e{red}^2 < 1$. However, those trajectories cannot reproduce the periodic features observed between 2005 and 2010. Therefore, we propose here a more detailed model of the microlensing curve with an accretion disk including a hotspot orbiting the central black hole. A Gaussian profile is added to a standard thin-disk profile to represent the hotspot, with its width fixed to 2 pixels full width half maximum. This corresponds to a physical size of $1.7 \times 10^{14}$ cm (0.07 light-days). The period, $P$, the luminosity ratio $L_{\mathrm{ratio}}=L_{\mathrm{disk}}/L_{\mathrm{hotspot}}$, the initial phase of the orbit $\theta_0$ and the accretion disk size $R_0$ are left as free parameters. The radius of the orbit is determined by the period and the black hole mass, which is kept fixed to $M\e{BH}=1.6 \times 10^{8} M_{\odot}$. To limit the number of free parameters, we assume that the hotspot is on a perfectly circular orbit, perpendicular to the plane of the sky in a face-on disk, but our results can be easily generalised to elliptical orbits and inclined accretion disks. We adopted flat priors: $P\e{o} \in [0,300]$\ days, $L_{\mathrm{ratio}} \in [0,10]$, $\theta_0 \in [0,2\pi],$ and $\log_{10}(R_0/\hbox{cm}) \in [14,17]$. We then computed the flux ratio along the pre-defined trajectory and compare it with the entire observed Euler light curve. We restrict our analysis to a smaller cutout of the magnification map, encompassing the best-fit trajectory, in order to keep the computational time manageable.

\begin{figure*}[h!]
    \centering
    \includegraphics[width=\textwidth]{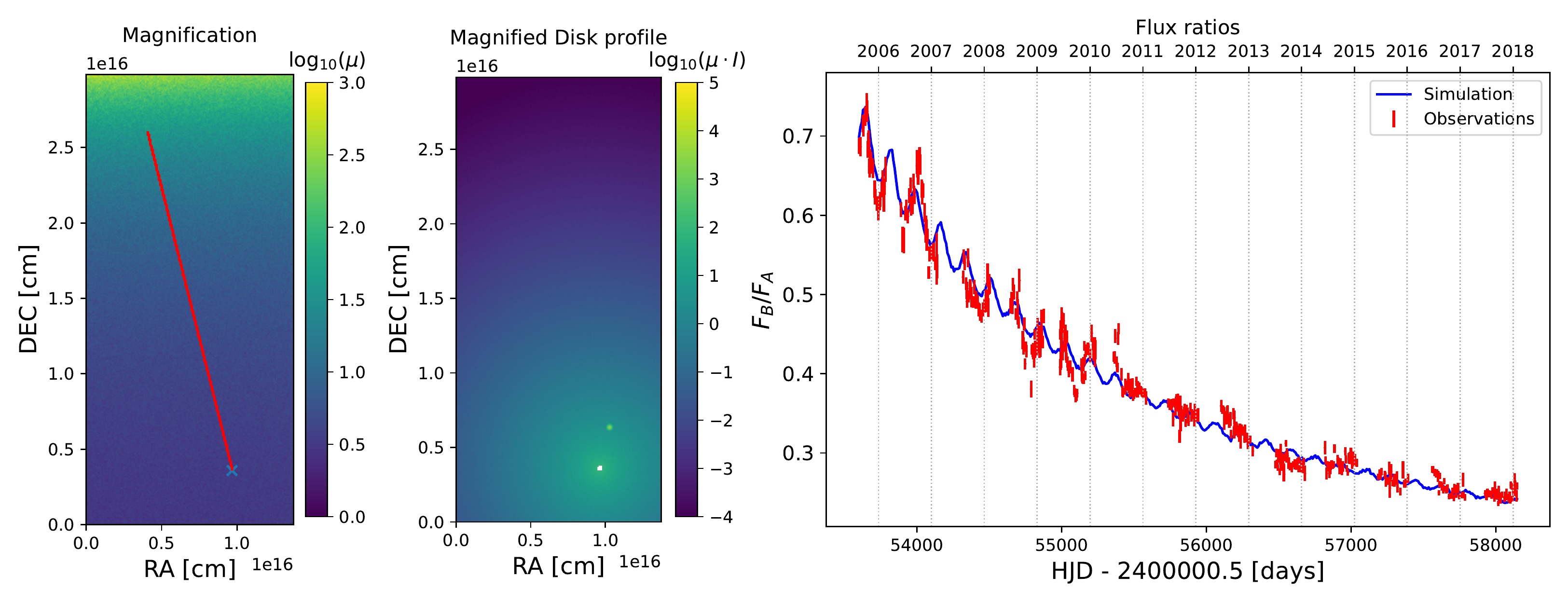}
    \caption{Inhomogeneous accretion disk simulations. \textit{Left panel:} Microlensing magnification map. The red line shows the trajectory that best fits the long-term microlensing trend (see Fig. \ref{fig:trend_fit}). \textit{Middle panel:} Microlensing magnified accretion disk profile, including a Gaussian hotspot orbiting the central black hole. The radius of the circular orbit is determined by the period, which is a free parameter, and the black hole mass, fixed to $1.6 \times 10^8 M_{\odot}$. \textit{Right panel:} Observed (red) and simulated (blue) flux ratios for the best-fit parameters. 
    An animated version of this figure is available at this \href{https://www.aanda.org/articles/aa/olm/2022/12/aa44440-22/aa44440-22.html}{link}.
    } 
    \label{fig:hotspot_simu}
\end{figure*}

\begin{figure}[h!]
    \centering
    \includegraphics[width=0.49\textwidth]{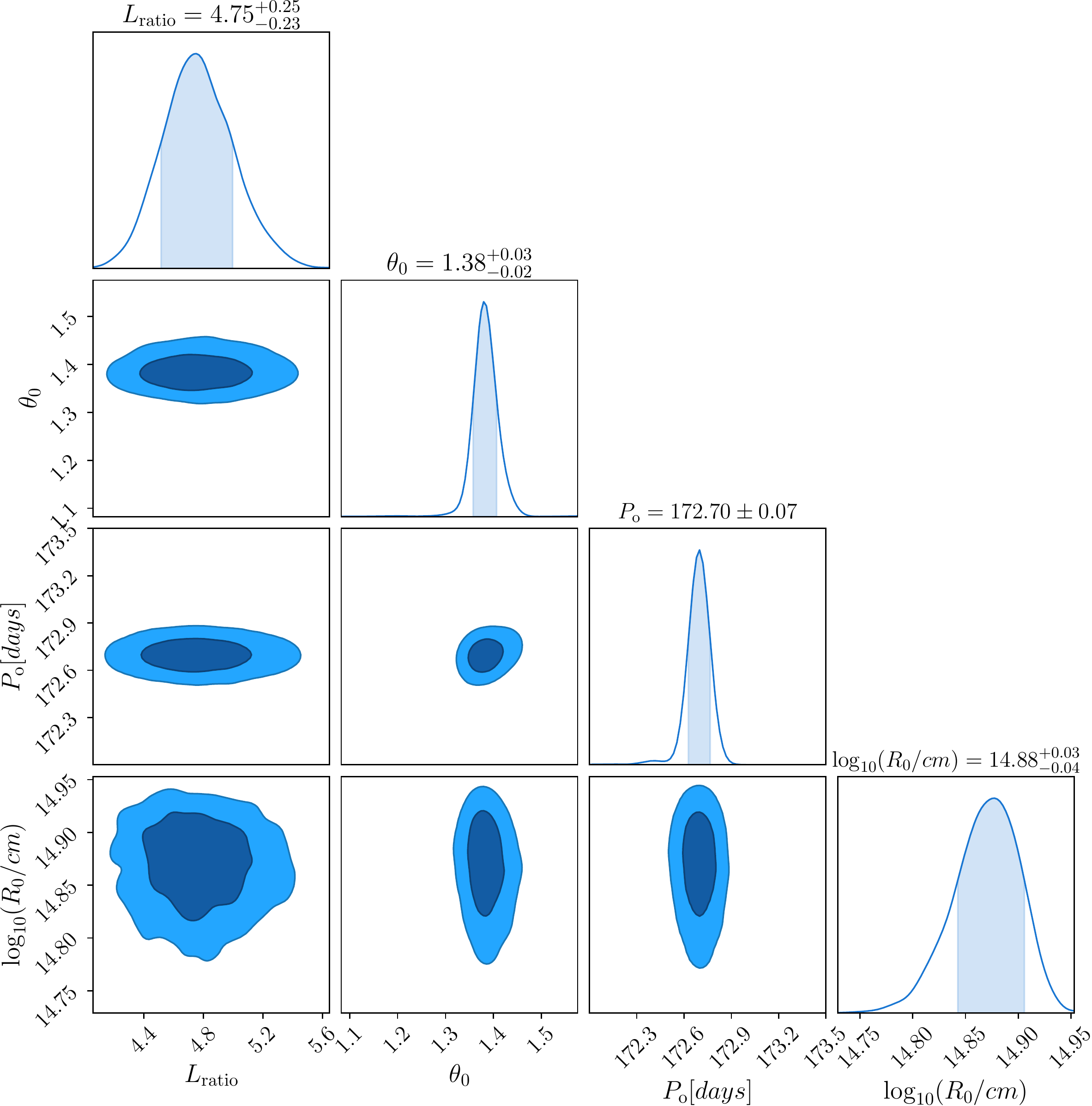}
    \caption{Posterior distributions of the model parameters. $L_{\mathrm{ratio}}$ is the luminosity ratio between the luminosity of the hotspot and the total luminosity of the disk, $\theta_0$ is the initial position angle of the hotspot, $P\e{o}$ is the orbital period, and $R_0$ is the accretion disk size. The mass of the black hole is kept fixed to $1.6\times 10^8 M_{\odot}$.}
    \label{fig:posterior_hotspotsimu}
\end{figure}

We used the Python nested sampling package \dynesty and the auto-differentiation package \jax \citep{jax2018} for the likelihood evaluation to make a Bayesian inference possible in a reasonable time on a single GPU. Figure \ref{fig:hotspot_simu} shows the trajectory in the magnification map, the microlensed accretion disk profile, and the best-fit simulated light curve. The posterior distributions of the free parameters are shown in Fig. \ref{fig:posterior_hotspotsimu}. 

 This model reproduces well the main features of the light curve but the amplitude of the oscillations is not always matched. Our goal is not to exactly reproduce the data as this would require finding the ensemble of tracks that are compatible and simultaneously account for the unknown complexity of the source. Considering the various hypotheses to be tested and the numerical complexity of such a fit, we do not attempt to perfectly model all the features seen in the microlensing curve. Still, with this simple physical model, we recover a similar value of the period ($P\e{o}=172.69^{+0.08}_{-0.06}$ days) as our purely empirical model presented in Sect. \ref{sec:model}. Interestingly, the luminosity ratio between the main accretion disk and the hotspot is constrained to $L_{\mathrm{ratio}}=4.8^{+0.2}_{-0.2}$. Although $R_0$ is left as a free parameter, we have an implicit prior on the accretion disk size coming from the pre-selected trajectory, chosen to fit the long-term trend of the microlensing curve. We nevertheless recover a similar luminosity ratio by selecting trajectories with 0.5$R_0$ and 2$R_0$ and repeating the experiment. 

\begin{figure}[h!]
    \centering
    \includegraphics[width=0.49\textwidth]{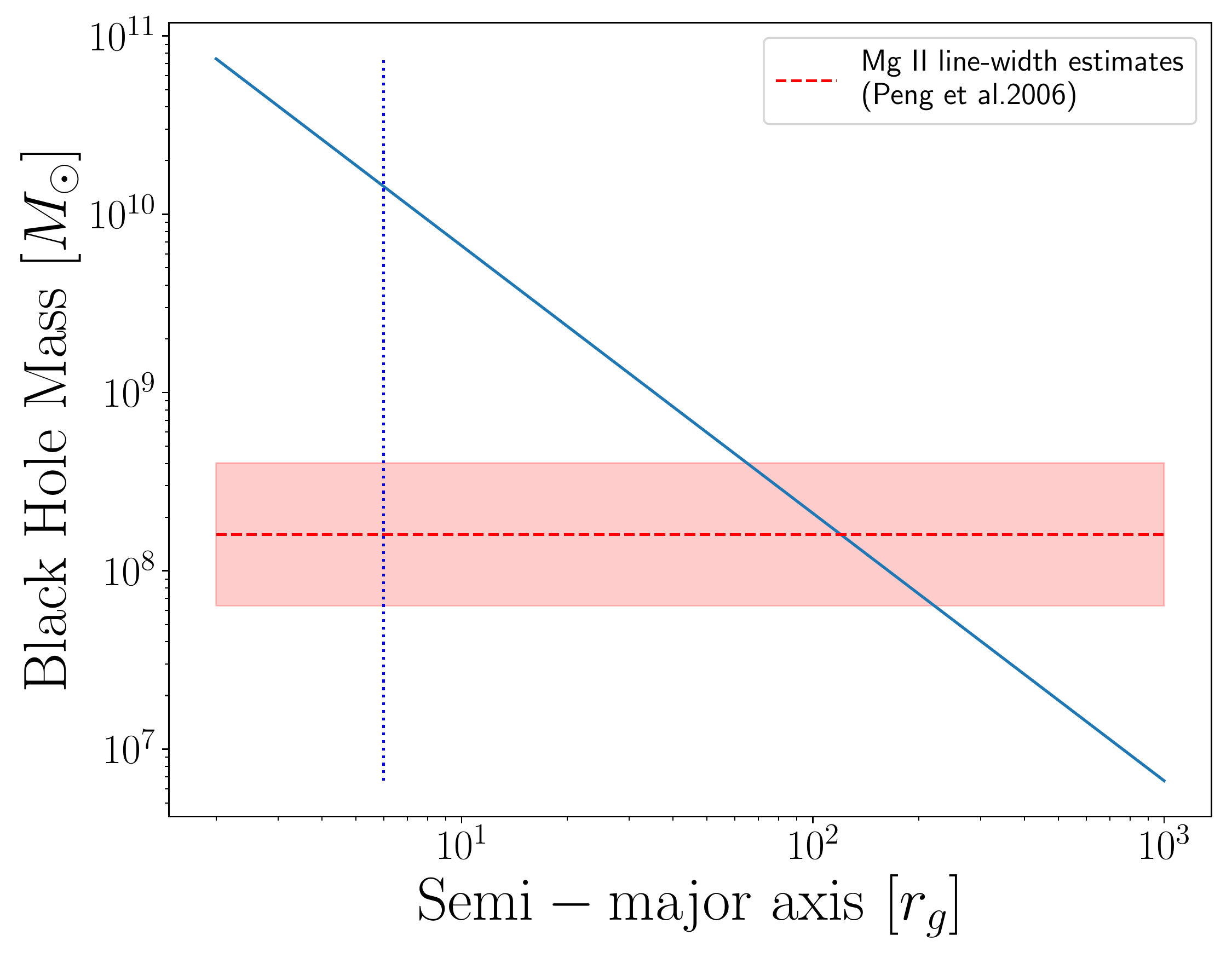}
    \caption{Black hole mass as a function of the semi-major axis of the hotspot's orbit. The orbital period is fixed to the observed one, $P_s = 75.4$ days. \cite{Peng2006a} estimates of the black hole mass imply that the emission region of the disk at the origin of the periodic signal is located at around 120 $r_g$ from the central black hole. The vertical blue line indicates the ISCO for a Schwarzschild black hole. }
    \label{fig:MBH-rISCO}
\end{figure}

The black hole mass estimates from \cite{Peng2006a} and the observed period constrain the distance of the hotspot from the central black hole, which is localised relatively far from the centre (120 $r_g$). If we remove our assumption on the black hole mass, we can compute the relation between the black hole mass and the semi-major axis of the orbit for a fixed period of 75.4 days. This is shown in Fig. \ref{fig:MBH-rISCO}. 

We have not yet discussed what could be the emission mechanism at play in this hypothetical hotspot, producing around 20\% of the total UV flux. A first explanation would be that it is powered by accretion onto a secondary smaller black hole, which would be similar to our second scenario. Alternatively, one can imagine that a compact region of the disk is significantly hotter than the rest of the accretion disk. To produce 20\% of the UV flux, this hotspot would preferably be located close to the ISCO, which would require the mass of the black hole to be largely underestimated. Bringing the hotspot to the ISCO would require an extremely large black hole mass of the order of $10^{10} M_{\odot}$. Although accretion disks are thought to be inhomogeneous at some level, the model of inhomogeneous accretion proposed by \cite{Dexter2011} rather produces small temperature fluctuations everywhere in the disk rather than in one single, hot, UV-emitting region. Models predicting numerous small inhomogeneities in the accretion disk would not produce the periodicity observed in the data. 

In addition, if the hotspot is powered by a mechanism other than accretion and is not bound by gravity, we would expect that the local inhomogeneity in the disk is rapidly disrupted by Keplerian shear. This is a similar argument made by \cite{Eracleous1995}, who estimated that inhomogeneities would dissipate on a timescale of order
\begin{equation}
    \tau\e{shear} \approx \frac{2rP_{\mathrm{Kep}}}{3h},
\end{equation}
where $r$ is the radius of the orbit, $P_{\mathrm{Kep}}$ is the Keplerian period and $h$ is the radial extent of the inhomogeneity. Considering that the inhomogeneity is approximately the size of the local height of the disk we obtain \citep{Veilleux1991, Eracleous1995}
\begin{equation}
    \tau\e{shear} \sim 7 \left( \frac{M\e{BH}}{10^8 M_{\odot}} \right)  \left( \frac{r}{100 r_g} \right) \left( \frac{T}{10^5 K } \right)^{-1/2}  yr,
\end{equation}
where $T$ is the local temperature of the disk.
This means that a hotspot localised at $\sim 120$ gravitational radii from a central black hole of $M\e{BH}=1.6 \times 10^8 M_{\odot}$ will dissipate in about 13 years because of Keplerian shear. This timescale is indeed longer than our observational baseline. The fact that such an extremely bright hotspot has appeared exactly during the high magnification event of image B would imply that they are much more common than expected. This is at odds with the observations of other lensed quasar microlensing events. The last possibility is that the same region of the disk is constantly heated, regenerating the hotspot on a timescale smaller than $\tau\e{shear}$. To our knowledge, no such mechanism is capable of producing one-fifth of the quasar luminosity on a single small region of the disk. In the same vein, obscuring a part of the disk (as proposed by e.g. \cite{Wyithe2002}) could introduce the observed periodicity but to match the amplitude it would require masking one quarter of the total quasar luminosity, and thus would also quickly be disrupted by Keplerian shear. 

Overall, accretion onto a secondary black hole seems the most plausible, if not the only, mechanism capable of producing the required amount of UV flux, while keeping the emission region sufficiently small to be significantly amplified by microlensing. 

\subsection{Precessing accretion disk}

Eccentric accretion disk models were originally proposed to account for asymmetric double-peaked emission lines seen in quasar spectra \citep{Eracleous1995, Strateva2003}. Two main mechanisms were put forward to explain their formation: (1) a perturbation of the disk by tidal forces induced by a smaller black hole and (2) elliptical accretion of the debris resulting from the disruption of a star. In the case of \Jzeroun, the black hole seems too massive to provoke a tidal disruption event. A star approaching the black hole would be swallowed without being disrupted \citep{Rees1988}. 

Regardless of the formation channel, eccentric disks precess due to the advance of the orbit's pericentre in a Schwarzschild potential. The precession angle per revolution is given by 
\begin{equation}
    \delta \phi = \frac{6\pi GM\e{BH}}{c^2 a(1-e^2)},
\end{equation}
where $a$ is the semi-major axis of the orbit and $e$ its eccentricity. 
This implies a precession period of \citep[e.g.][]{Eracleous1995, Storchi-Bergmann2003}
\begin{equation}
    \label{eq:precession_period}
    P\e{prec} \sim 3.29 \left( \frac{M\e{BH}}{10^8 M_{\odot}} \right) \left( \frac{r}{100 r_g} \right)^{5/2} \,\hbox{yr}, 
\end{equation}
where $r$ is the pericentre distance of the orbit.
Fixing $P\e{prec}$ to the observed period in the source plane, $P_s = 75$ days, we obtain from Eq.\,\eqref{eq:precession_period} an estimate of the orbit radius of the emitting region in precession: 
\begin{equation}
    r\e{prec} \sim 27 r_g. 
\end{equation}

Following the argument of \cite{Eracleous1995}, the timescale on which an orbit will circularise due to differential precession can be estimated as
\begin{equation}
    \tau\e{circ} \sim 2\times 10^4 \left( \frac{M\e{BH}}{10^8 M_{\odot}} \right)  \left( \frac{r}{100 r_g} \right)^2 (1+e) \left( \frac{T}{10^5 K } \right)^{-1/2} \,\hbox{yr}. 
\end{equation}
If we assume a local temperature of the UV emission region of $\sim$ 5000K, the circularisation timescale at $r\e{prec}$ is $\tau\e{circ} \sim$\ 2600 yr. This timescale is to be compared with the local viscous timescale to determine if the disk can remain eccentric for a sufficiently long time. The local viscous timescale is given by \citep{Frank1992}:
\begin{equation}
\tau\e{visc} \sim 3150 \ \alpha^{-4/5} \dot{m}^{-3/10} \left( \frac{M\e{BH}}{10^8 M_{\odot}} \right)^{3/2}  \left( \frac{r}{100 r_g} \right)^{5/4}  \,\hbox{yr}, 
\end{equation}
where $\alpha$ is the viscosity parameter as defined by \cite{Shakura1973}, and $\dot{m}$ is the accretion rate in $M_{\odot}\cdot$yr$^{-1}$. Assuming a typical value for $\alpha$ of 0.2 and a typical accretion rate of 1$M_{\odot}\cdot$yr$^{-1}$, the viscous timescale of the precessing region can be roughly estimated to $\tau\e{visc} \sim$\ 4500 yr. The two timescales are comparable, which indicates that the differential precession plays an important role in the circularisation of the disk. As discussed in \cite{Eracleous1995}, only the outer part of the disk ($ r>100r_g$) could maintain a significant eccentricity. This would, however, lead to a much longer precession period, of the order of $\sim$ 1000 years, hence impossible to detect. This scenario seems therefore improbable unless the eccentricity has developed recently or is maintained by tidal forces.

We finally propose that a detached disk is in precession; not because of its eccentricity but because it would be subject to \cite{Lense1918} differential torques if the disk is not aligned with the black hole spin \citep{Bardeen1975}. The disk might break into several rings, which precess at different rates \citep{Nixon2012, Nealon2015}. In this scenario, the orientation of the disk relative to the line of sight might change periodically, hence modulating the luminosity. Alternatively, the detached disk might also shadow periodically the central source. The precession period is given by 

\begin{equation}
    P_{LT} = \pi \frac{c^3 a^3 (1 - e^2)^{3/2}}{G^2M\e{BH}^2 \chi},
\end{equation}where $\chi$ is the dimensionless black hole spin parameter. Assuming no eccentricity, and black hole spin in the range $\chi = 0.1 - 0.9$, we estimate the radius of the detached ring in the range $8 - 17$ $r_g$ to match the observed period, assuming that the modulation of the signal occurs at twice the precession frequency. This result is difficult to accommodate with theoretical expectations, which predict that the inner disk ($r \lesssim 100 r_g$) should align rapidly with the black hole spin because of the differential \cite{Lense1918} torques \citep[see e.g.][and references therein]{Natarajan1998, King2005, Nixon2012}. However, recent general-relativistic magnetohydrodynamic simulations by \cite{Liska2021} have shown that the alignment radius might be as small as 5 to 10 gravitational radii in the case of thin, highly tilted disks around rapidly rotating black holes. Although it cannot be completely excluded, this scenario of an accretion disk in rapid precession would face a second difficulty; if a detached disk is obscuring periodically the central source, it should also leave an imprint in image A, which is not observed. We did not detect any significant power in the Lomb-Scargle periodogram of image A at this frequency. 
Even in the case of a strong micro-magnification gradient across the accretion disk in image B, it is difficult to imagine a configuration where a detached precessing disk would absorb up to 20\% of the flux of image B while staying unnoticed in image A. 

\section{Discussion}
\label{sec:discussion}

\Jzeroun has now been monitored in the optical for 15 years, thus allowing us to obtain a robust measurement of the time delay. Given the time delay, it is now obvious that image B has encountered a high magnification event during the period 2003-2008, with a possible caustic crossing between 2003 and 2006. The quasar has moved away from the caustic and is now de-magnified by microlensing. Over the period 2005-2018, the long-term microlensing trend is typical of a system exiting a micro-caustic and is well reproduced in our simulation (see Fig. \ref{fig:trend_fit}). It is however not clear why the microlensing amplitude has a maximum in the middle of the season 2003, decreases in 2004 and reaches a second maximum of similar amplitude in 2006. It is possible that this double peak is the signature of double caustic crossing with the quasar entering and exiting the caustic 2 years apart. 
However, we could not find any trajectories matching both the long-term trend over the period 2005-2018 and the double peak in 2003 and 2005 for a single disk size. We speculate that a more complex source-size effect plays an important role during the high magnification event between 2003 and 2005, and a simple thin disk model following a rectilinear trajectory through the microlensing magnification pattern is not sufficient to represent our data. 

Thus, for the rest of the analysis, we focused only on the Euler data, covering the period 2005-2018, which contain another key feature of the microlensing curve; periodic oscillations of the image flux ratio. These oscillations are detected at a period of 172.6 days during the high magnification event. The amplitude of this signal decreases concomitantly with the microlensing magnification of image B. We note that this period of 172.6 days also matches with the peak observed in the first season in 2003, providing supporting evidence that this peak has the same physical origin as the rest of the oscillations. This feature in the first season of the SMARTS data corroborates the hypothesis that the periodic signal originates from a sub-structure (possibly a secondary black hole) orbiting the quasar.

The scenario of a binary black hole as the origin of this periodicity is appealing as it naturally explains both the amplitude and period of the signal, whether or not the secondary black hole has its own accretion disk and associated UV emission. Our simple model shows that the motion of the main accretion disk in the source plane around the centre of mass of the system would be sufficient to reproduce the observed signal with a modest mass ratio ($q \lesssim 10$), commonly observed in numerical simulations \citep[e.g.][]{Volonteri2003,Volonteri2009}. Similarly, if the secondary black hole has its own light emission and its orbit moves it periodically into higher and lower magnification regions of the microlensing map, only a modest luminosity ratio of $\sim$ 5 is required to fit our data. Assuming a rough scaling relation $M\propto L^{0.7}$ \citep{Woo2002}, this leads to a mass ratio of $q\sim3$, similar to that expected from a companion causing oscillations of the primary disk. These numbers are derived under simplifying assumptions: that the quasar disk is seen face-on and the orbital motion is circular and perpendicular to the line of sight. Including all the orbital degrees of freedom might be necessary to fully explain the shape of the oscillation seen in the microlensing light curve but this is left for future work. We also assumed that the light profile of the quasar follows a simple thin-disk model or a Gaussian profile. Including a more realistic light profile of the interacting accretion disk might also better reproduce the observed data. Finally, there is a possibility that the two black holes have similar UV brightness. We did not fit the mass and luminosity ratios at the same time because these two quantities are degenerate and it did not allow us to obtain meaningful constraints. However, in this scenario, the microlensing light curve is modulated at half the orbital period. This would only change the orbital parameters marginally ($r$=$1.6 \times 10^{-3}$ pc) but leave the rest of our conclusions unchanged. This last possibility would qualitatively explain the second harmonic peak at 342 days seen in the periodogram, although it could also be explained by the amplitude of the signal decreasing over time, which would artificially create power at higher harmonics as well.

The main issue of the SMBBH scenario is the short lifetime of these systems, due to the rapid decay of their orbit through GW emission. This problem is in fact far from being insurmountable. Several simulations \citep{Tang2017, Moody2019, Munoz2019, Munoz2020, Bortolas2021} have demonstrated that, in some cases, the torques induced by circumbinary disks could counteract the GW-induced torques and slow down the decay of the orbit \citep[see e.g. Sect. 2.2.2.2 of ][for a review of this issue]{LisaCollaboration2022}. This mechanism could delay significantly the merger of the two black holes, or even cause the two black holes to out-spiral. In this case, close binaries separated by a few hundred $r_g$ to a thousand $r_g$ would be much more common.

Finally, we estimate the characteristic strain and frequency of the GW that could be observed from this system and compare with the sensitivity curves of current and future PTA experiments. For simplicity, we treat the two images separately and ignore 
interferences although these can be used to break the mass-sheet degeneracy \citep{Cremonese:2021puh}. This is enough to get an order of magnitude of the characteristic strain. The GW frequency, $f\e{gw}$, corresponds to twice the observed frequency. We assume the latter to match the orbital frequency, such that \,$f\e{gw}=2/P\e{o} =1.34 \cdot 10^{-7}$ Hz, which falls in the PTA band $[10^{-9}$, $10^{-6}]$ Hz. We note that this corresponds to a wavelength of $\lambda\e{gw} = 0.072$ pc, which is comparable to the Schwarzschild radius of the lens galaxy, for which wave effects \citep{Caliskan:2022hbu} and polarisation distortions \citep{Dalang:2022} can be at play.\footnote{Since the wavelength is of the order of the Schwarzschild radius of the lens galaxy and therefore much larger than the Schwarzschild radius of the microlens, we do not expect microlensing to affect the GW.}

\begin{figure*}
    \centering
    \includegraphics[width=\textwidth]{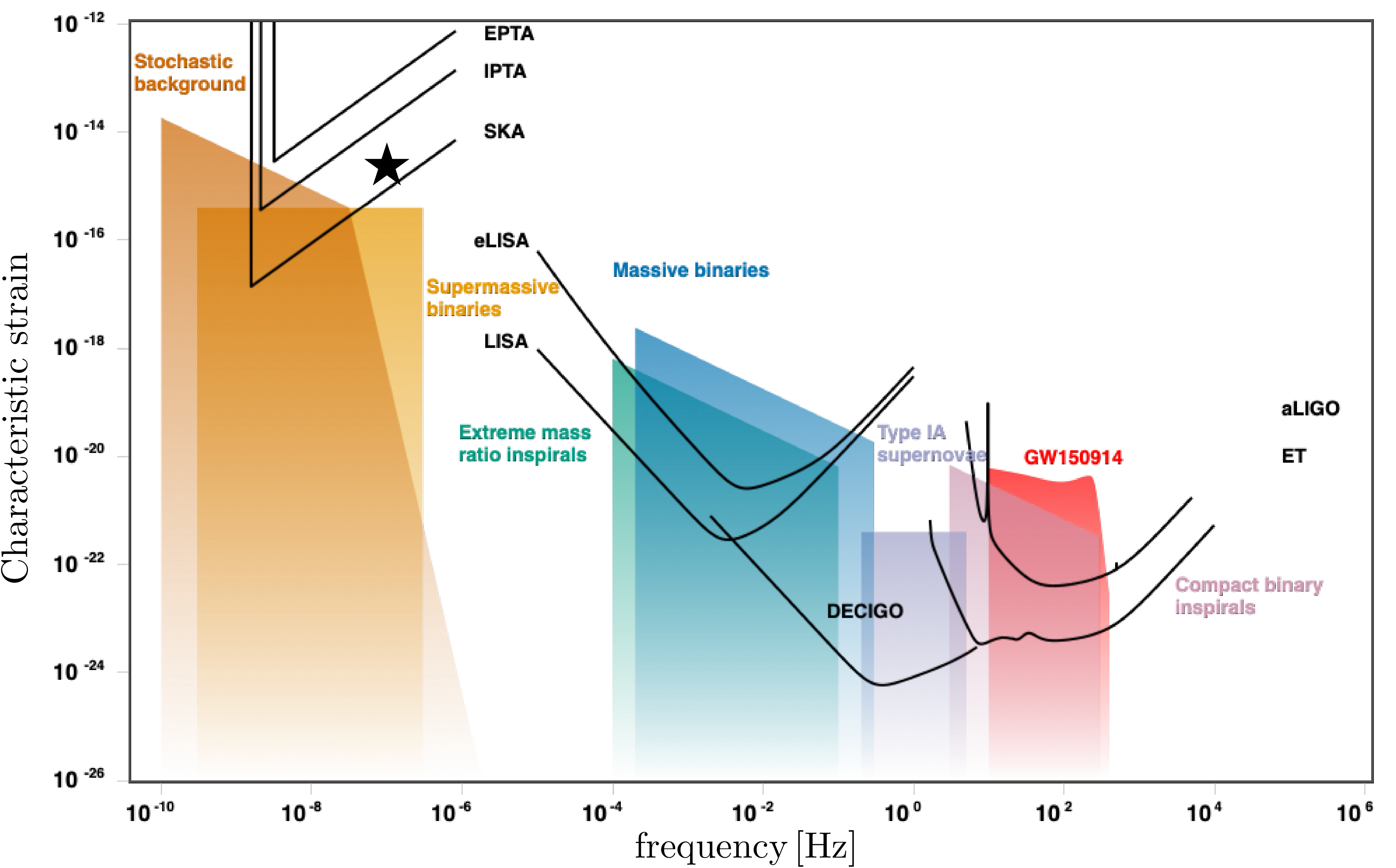}
    \caption{Adapted from \cite{Moore:2014lga}. The approximate characteristic strain of the GW signal is depicted by the star at $f\e{gw}\simeq 1.34 \cdot 10^{-7}$ [Hz], falling in the PTA band [$10^{-9}$, $10^{-6}$] [Hz], and lying above the approximate sensitivity curve of SKA but below that of European PTA (EPTA) and International PTA(IPTA). The detectability with SKA will depend on the exact details of the pulsar array.}
    \label{fig:strain}
\end{figure*}

We assume the total mass of the system to be fixed $M\e{tot} = M_1+M_2= 1.6 \cdot 10^{8} M_\odot$ and let the mass ratio $q\equiv M_1/M_2\in [1,10]$ such that $M_2(q) = M\e{tot}/(1+q)$ and $M_1(q) = q \cdot M_2(q)$. The time evolution of the observed GW frequency for a binary system in quasi-circular orbit that is slowly losing energy to GWs is given by \cite[see e.g.][]{Maggiore:1900zz}
\begin{align}
\dot{f}\e{gw}(q)= \frac{96}{5} \pi^{8/3} \l( \frac{G\mathcal{M}_z(q)}{c^3}\r)^{5/3} f\e{gw}^{11/3}\,,
\end{align}
where the `{redshifted chirp mass'} is defined as 
\begin{align}
\mathcal{M}_z(q) =(1+z) \frac{[M_1(q)M_2(q)]^{3/5}}{(M_1(q)+M_2(q))^{1/5}}\,.
\end{align}
A PTA experiment is sensitive to a linear combination of the plus and cross polarisations $h(t)=F_+ h_+(t) + F_\times h_\times(t)$, where the factors $F_+$ and $F_\times$ are combinations of trigonometric functions that depend on the geometry of the pulsar array and satisfy $F_+, F_\times \in [-1,1]$ \citep{Moore:2014lga}. The two independent and magnified polarisations of a GW emitted by a binary system in quasi-circular orbit, expressed in terms of observed quantities, read \citep{Maggiore:1900zz,1992Schneider}
\begin{align}
h_+(t) &= \sqrt{|\mu|} \mathcal{A}\e{o}(q) \l(\frac{1+\cos^2(\iota)}{2}\r) \cos[\Phi(t)]\,\\
h_\times(t) & = \sqrt{|\mu|} \mathcal{A}\e{o}(q) \cos \iota \sin[\Phi(t)],\, 
\end{align}
where $\cos \iota = \boldsymbol{\hat{L}} \cdot \boldsymbol{\hat{n}}$ is the cosine of the angle between the orbital plane $\boldsymbol{\hat{L}}$ and the line of sight $\boldsymbol{\hat{n}}$, $\mu$ is the magnification of the considered image, $\Phi(t)$ is the phase of the GW and $\mathcal{A}\e{o}(q)$ is the amplitude of the unlensed GW at the observer, which reads
\begin{align}
\mathcal{A}\e{o}(q) = \frac{4\l( \frac{G \mathcal{M}_z(q)}{c^2}\r)^{5/3} \l(\frac{\pi f\e{gw}}{c}\r)^{2/3}}{D\e{L}(z\e{s})}\,.
\end{align}
The luminosity distance at redshift $z\e{s}=1.29$ can be computed using a flat $\Lambda$CDM cosmology and we find $D\e{L}(z\e{s})= 9.3$ Gpc. The (squared) characteristic strain is defined as \citep{Moore:2014lga}
\begin{align}
[h\e{c}(f)]^2 \equiv 4 f^2 \l|\t{h}(f) \r|^2\,.
\end{align}
For a binary in quasi-circular motion, the GW is nearly monochromatic, such that the Fourier transform $\t{h}(f)$ of $h(t)$ can be computed using a saddle point approximation \citep{PhysRevD.62.124021, Moore:2014lga} and we find 
\begin{align}
\t{h}(f) = \sqrt{\frac{|\mu|}{\dot{f}}} \frac{\mathcal{A}\e{o}(q)}{2}\l( F_+ \frac{1+\cos^2(\iota)}{2} + F_\times \cos(\iota) \r)\,,
\end{align}
where each term inside the brackets belongs to the interval $[-1,1]$. Therefore, we estimate the characteristic strain at frequency $f\e{gw}$ and mass ratio $q$ to be of order
\begin{align}
h\e{c}(f\e{gw},q) \sim \sqrt{\frac{f\e{gw}^2 |\mu|}{\dot{f}\e{gw}}} \mathcal{A}\e{o}(q)\,.
\end{align}
For image $A$ with lensing macro-magnification $\mu_A = 2.27$, this implies $h\e{c}(f\e{o},q) \in [1.4 \cdot 10^{-15}, 2.5 \cdot 10^{-15}]$ for $q\in[1,10]$, which is above the approximate sensitivity curve of the Square Kilometre Array (SKA) and below that of the European Pulsar Timing Array (EPTA). Here, we chose a low estimate of the lensing magnification $\mu_A$ from the models of \cite{Morgan2008} but the lensing magnification could be up to five times larger if the stellar mass fraction is rather of the order of $f\e{M/L} = 0.3$ instead of 0.9. The exact details of the pulsar array will be needed to estimate if the signal is observable with the SKA. This prevents a more precise conclusion on the detectability of this GW signal. Figure \ref{fig:strain} shows how the frequency and the characteristic strain of the signal compare with estimated sensitivity curves of the current and future PTA experiments. 

\section{Summary and conclusion}
\label{sec:conclusion}
We report the first detection of periodic oscillations in the flux ratios of multiple images of a lensed quasar. These oscillations are visible in the microlensing curve of \Jzeroun over the period 2003 - 2010, corresponding to a high magnification event of image B. Their amplitude decreases as image B is less and less magnified by microlensing. We measure from a simple sinusoidal model a period of $172.6 \pm 0.9$ days in the observer frame, corresponding to $75.4 \pm 0.4 $ days in the quasar rest frame. The same period is detected in each of the Euler and SMARTS microlensing curves as well as in the joint Euler and SMARTS curve. This period is also confirmed by the Lomb-Scargle analysis, with a large peak at 171 days.

From these observations, we have developed several hypotheses to explain the origin of this periodicity. We rank our hypotheses from the most probable to the least probable: 

\begin{enumerate}
    \item \Jzeroun hosts a SMBBH: we have demonstrated from a simple model that a binary black hole naturally reproduces both the amplitude and period of the observed signal if the mass ratio is of the order $q\lesssim 10$.  Assuming the black hole mass estimate from \cite{Peng2006a}, we derive a coalescence time due to GW emission of $\sim$ 1000 years, extremely short compared to the age of the quasar. However, the transfer of angular momentum from a circumbinary disk to the binary system could significantly increase the lifetime of such close binaries, making them much more likely to be observed. 
    
    \item The accretion disk contains a large inhomogeneity: this scenario also fits our observations but it requires one-fifth of the total UV luminosity to be emitted by a compact, hotter region of the disk. If not bound by gravity, this scenario also faces the problem of Keplerian shear, which would disrupt the hotter region of the accretion disk on short timescales. Accretion seems the only plausible mechanism to produce such an amount of UV flux over a sufficiently compact region to be microlensed. If the hotspot in the disk is powered by accretion, then this hypothesis is similar to our first scenario.    
    
    \item The accretion disk is in precession: the short period of the signal means that the inner part ($r<30r_g$) of the disk must be in precession. In this scenario the disk would be subject to a strong differential precession, leading to a rapid circularisation of the orbit in the case of an eccentric disk, or to a rapid alignment of the accretion disk with the black hole spin, in the case of Lense-Thirring precession.
    
    \item Microlensing by binary stars: this last scenario is ruled out by the small separation between the stars that is imposed by the observed period. A pair of 1$M_{\odot}$ would need to be separated by 1.01 AU to produce the observed period. Such small separations in the lens plane only produce an extremely small motion of the micro-caustic in the source plane, making it impossible to reproduce the amplitude of the observed signal.
    
\end{enumerate}

In the absence of the zoom-in effect produced by microlensing, these oscillations will likely no longer be observed in photometric light curves, but they might reappear if a high magnification microlensing event reoccurs in either of the two images. Over the ten years of the Rubin Observatory's Legacy Survey of Space and Time, it is likely that this system will again approach a microlensing caustic, opening the possibility to trigger spectroscopic follow-up to confirm or rule out different scenarios. Even in the absence of a strong microlensing magnification, a periodic change in the emission lines' profiles might still be detectable.

Finally, the best way to confirm the presence of a SMBBH might very well be the detection of GWs emitted by this source. We show that the mass of this system should be sufficient to be above the noise level of upcoming PTA experiments. This is speculative for the moment, but it might be possible, in the future, to obtain an extremely precise measurement of the time delay from the GW signal, with strong implications for cosmology. This system might be an extraordinary laboratory to test Einstein's theory of general relativity at the crossroad of two of its most famous predictions: the gravitational lensing effect and the propagation of GWs.

\begin{acknowledgements}
The authors would like to thank all the Euler observers as well as the technical staff of the Euler Swiss telescope who made the uninterrupted observation over 15 years possible. M.M. warmly thanks Marta Volonteri, Paul Schechter, Lucio Mayer and Georgios Vernardos for useful comments and informative discussions. COSMOGRAIL is supported by the Swiss National Science Foundation (SNSF) and by the European Research Council (ERC) under the European Union’s Horizon 2020 research and innovation programme (COSMICLENS: grant agreement No 787886). M.M.\,acknowledges support by the SNSF through grant P500PT\_203114.  This project was conceived during the Lensing Odyssey 2021 conference. This research made use of Astropy, a community-developed core Python package for Astronomy \citep{Astropy2013, Astropy2018}, the MCMC chain plotting package ChainConsumer \citep{Hinton2016}, and the 2D graphics environment Matplotlib \citep{Hunter2007}. 

\end{acknowledgements}

\bibliographystyle{aa}
\bibliography{biblio}

\begin{appendix}
\section{Reverberated signal in the microlensing curve}\label{Appendix:A}
\begin{figure}
        \centering
        \includegraphics[width=0.55\textwidth]{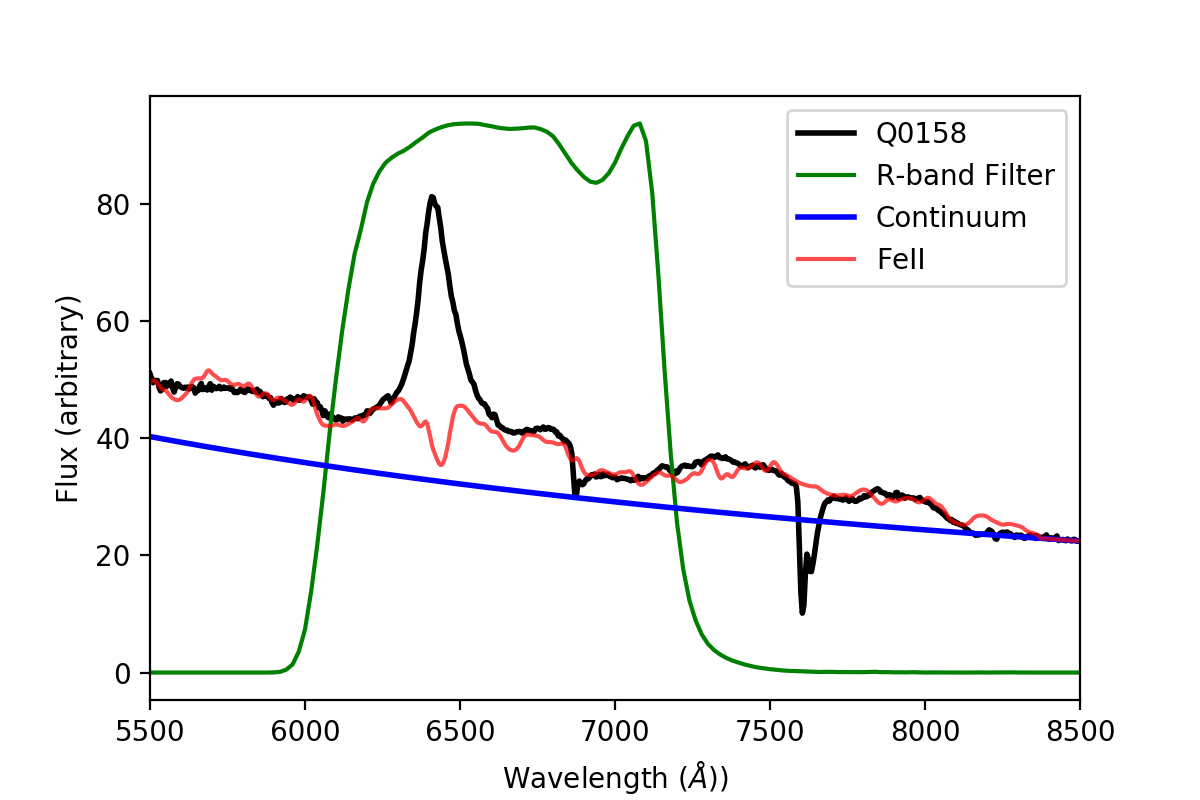}
        \caption{Spectrum of image A of \Jzeroun (black). The blue and red curve correspond to the best fitted model of continuum and \ion{Fe}{ii} emission. The green curve shows the transmission curve of the Euler $R-$band filter. }
        \label{fig:MgII}
\end{figure}

\begin{figure}
        \centering
        \includegraphics[width=0.5\textwidth]{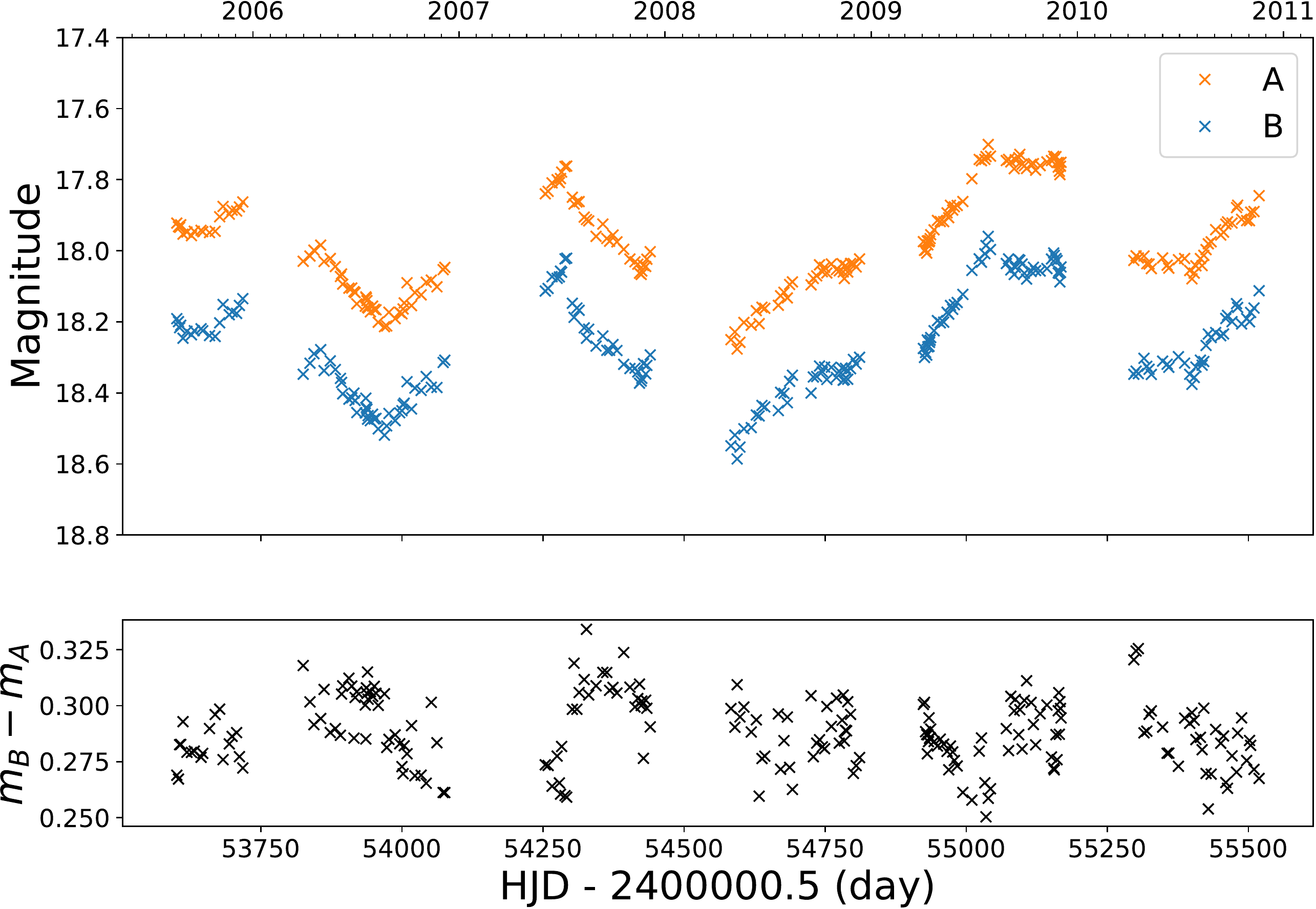}
        \caption{Simulated light curves generated from a DRW realisation, including the echoed signal of the BLR (top panel) and difference light curve between image B and image A (bottom panel). }
        \label{fig:DRWsim}
\end{figure}

\cite{Sluse2014} have suggested that, in the presence of microlensing, a deformed imprint of the intrinsic variability signal could appear in the difference light curve of a pair of lensed images because emission arising from differently microlensed regions are mixed in a given observing filter. The two main sources of differentially microlensed emission present in the $R$ band are the power-law continuum emission, and the emission arising from the broad emission lines. The continuum emission region is smaller than a microlens Einstein radius, and is therefore most prone to microlensing, while the emission from the broad line is much less microlensed. As explained in \cite{Paic2021}, the spectra of \Jzeroun observed by \cite{Faure2009} indicate that $\sim 40$\% of the $R-$band flux arises from the BLR (Fig.~\ref{fig:MgII}). Based on this estimate of the fraction of non-microlensed flux in the $R$ band, we generated mock light curves of the lensed images and evaluated the amplitude of the flickering introduced by the above effect. Following \cite{Paic2021}, we emulated the continuum signal $F_c$ using a DRW model \citep{Kelly2009, Macleod2010}, and added to it a reverberated BLR signal responding to the intrinsic variations with a lag of $\tau = 65$ days through a top-hat transfer function $\Psi(t, \tau)$. Following this procedure, the flux of image $i$ (i.e. $A$ or $B$), already corrected from the cosmological time delay, can be written as 

\begin{equation}
        F_i(t) = M_i \mu_i(t) F_{c}(t) + M_i f_{BLR} [\psi(t, \tau) \ast F_c(t)],
\end{equation}where $M_i$ is the absolute value of the macro-magnification of the image, $\mu_i(t)$ the time variable microlensing magnification and $f_{BLR}$ is the fraction of reverberated flux. We consider a constant microlensing amplification of image B fixed to the maximal micro-magnification observed in 2005 (i.e. $\mu_B (t) = 2),$ and we assume that A is unaffected by microlensing by fixing $\mu_A (t) = 1$. We also fix the non-microlensed flux ratio to the macro model prediction, $M_B/M_A = 0.44$. We use this physically motivated model to generate 5000 microlensing curves from different DRW realisations, with the same sampling and the same photometric noise as the real data. The mean flux level of $F_c(t)$ is arbitrarily fixed to 100 and the DRW timescale parameter, $\tau \e{DRW} =817 $ days, is obtained by fitting the light curve of image A with JAVELIN \citep{Zu2013}. The amplitude of the DRW,  $\sigma \e{DRW} = 20$ (in flux units), is adjusted so that the variations of the total (i.e. reverberated + continuum) flux in image A matches the observed variations. An example of light curves generated from this model is shown in Fig. \ref{fig:DRWsim}. 

First, we find a maximum peak to peak amplitude of the flickering of, at most, 0.10 mag. This corresponds to half of the observed amplitude of the observed periodic signal. We note that the detailed structure of the BLR signal does not matter much. For instance, a similar signal is observed if we assume a constant BLR contribution with time. The scattered emission from the continuum \citep[e.g.][]{Sluse2015, Hutsemekers2020}, would produce a similar effect as long as it arises from a region large enough to remain non microlensed. This simulation shows that, even under conservative assumptions, the signal arising from a larger region than the continuum, may produce red-noise with too low amplitude to mimic the observed signal.

\begin{figure*}[h!]
    \centering
    \includegraphics[width = 0.8 \textwidth]{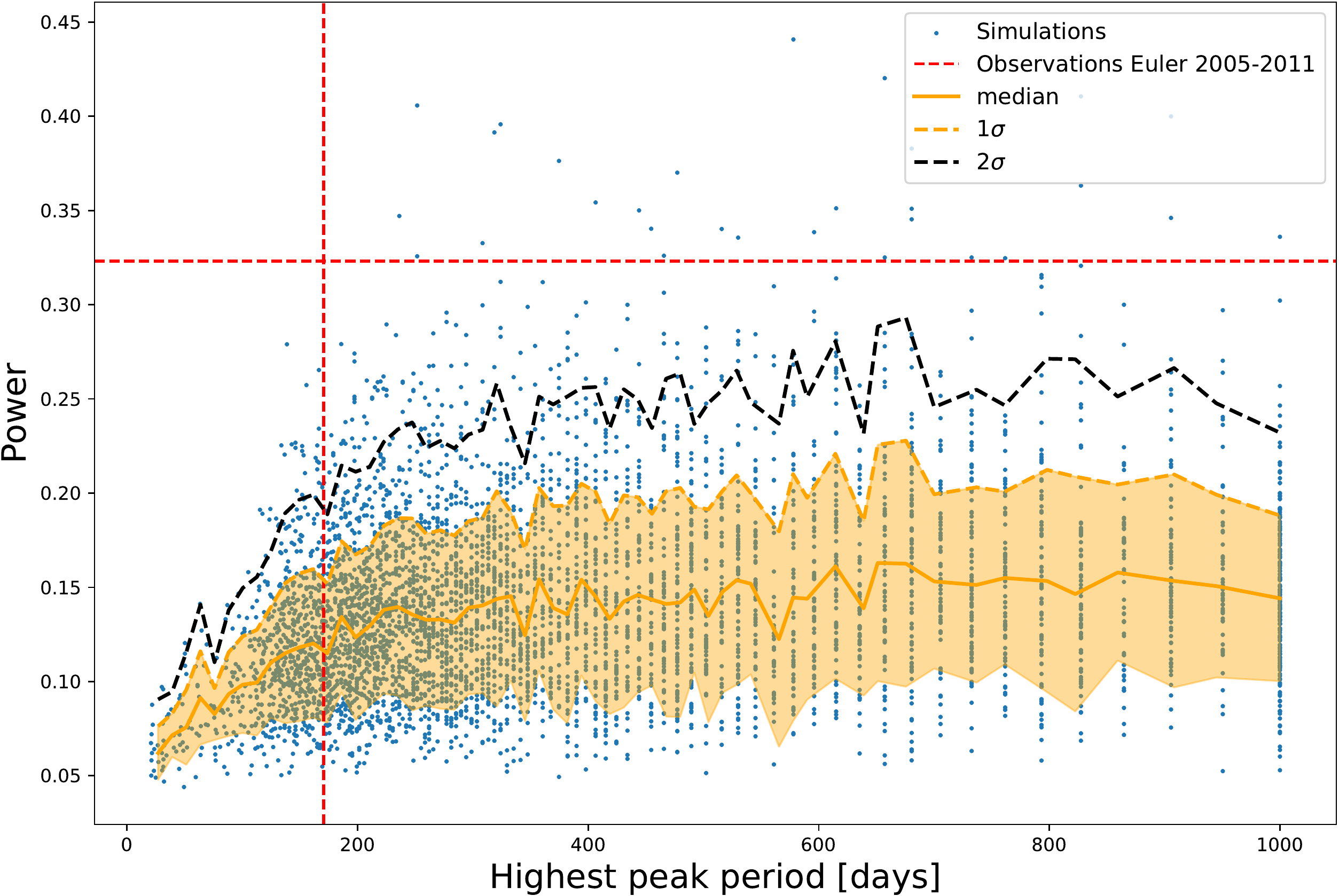}
    \caption{\label{fig:significance} Power of the highest peak in the GLS periodogram as a function of the peak period for 5000 simulated microlensing curves. The highest peak power and the period observed in the Euler data over the period 2005-2011 are indicated as dashed red lines. The 1$\sigma$ (dashed orange line) and 2$\sigma$ level (dashed black line) are computed in 80 different period bins of width 18.5 days.}
\end{figure*}

Second, we compute the GLS periodogram over a frequency range $[20^{-1} - 1000^{-1}]\ \mathrm{days}^{-1}$ for each of the simulated light curves and compare the power of the highest peak in the periodogram with that of the observed data. Here, we restrict our analysis over the period 2005-2011, where the periodic signal is clearly seen in the data. The results of this test are shown on Fig. \ref{fig:significance}. 

As discussed in \cite{Oneil2022}, peaks at any frequencies should be considered since we have no a priori reason to select the particular frequency observed in the real data. We thus conclude from these simulations that a spurious detection of the periodicity due to the intrinsic variability of the quasar echoed in the microlensing curve is rejected at 99.4\% confidence level (3.7$\sigma$).

Finally, we test alternative micro and macro-magnification models, selected to match approximately the minimum magnitude difference between image A and image B, $m_B - m_a \sim 0.3 $mag, as observed in 2005. These models are rejected with a significance ranging from 1.9 to 6$\sigma$ as summarised in Table \ref{tab:alternative_model}. We also consider different sizes of the BLR by varying the lag $\tau$ from 35 to 130 days. Although the model with a longer lag can only be excluded at 1.2$\sigma$ when all light curves are considered, this model cannot explain the short observed period as none of the simulated curves with a highest peak period below 200 days have more power at these frequencies than the observed data. Considering only the curves with a highest peak period within the range 165-175 days, this model is excluded at 7.2$\sigma$.

\begin{table*}[h!]
\centering
\renewcommand{\arraystretch}{1.5}

\caption{\label{tab:alternative_model} Rejection significance, $\sigma \e{r}$, for alternative magnification and reverberation models. Model parameters, $f_{BLR}$, $\tau$, $M_A$, $M_B$, $\mu_A$, and $\mu_B$, are defined in Appendix \ref{Appendix:A}. They are selected to match approximately the minimal magnitude difference, $m_B - m_A$, observed in 2005. The last column detail the rejection significance $\sigma \e{r, 165-175}$ when considering only the simulated light curves with highest peak period within the range 165-175 days.}

\begin{tabular}{l|llllll|c|cc}
Model & \multicolumn{1}{c}{$f_{BLR}$} & \multicolumn{1}{c}{\begin{tabular}[c]{@{}c@{}}$\tau$\\ $[$days$]$\end{tabular}} & \multicolumn{1}{c}{$M_A$} & \multicolumn{1}{c}{$M_B$} & \multicolumn{1}{c}{$\mu_A$} & \multicolumn{1}{c|}{$\mu_B$} & \begin{tabular}[c]{@{}c@{}}$m_B - m_A$ \\ $[$mag$]$\end{tabular} & $\sigma \e{r}$ & $\sigma \e{r, 165-175}$ \\ \hline
fiducial & 0.4 & 65 & 2.26 & 1.01 & 1.0 & 2.0 & 0.29 & 3.7$\sigma$ & 5.2$\sigma$ \\
alternative micro-model & 0.4 & 65 & 2.26 & 1.01 & 0.5 & 1.1 & 0.32 & 2.4$\sigma$ & 3.3$\sigma$ \\
alternative micro-model & 0.4 & 65 & 2.26 & 1.01 & 1.5 & 2.8 & 0.34 & 6.0$\sigma$ & 8.0$\sigma$ \\
alternative macro-model & 0.4 & 65 & 5.00 & 1.01 & 1.0 & 4.8 & 0.31 & 1.9$\sigma$ & 2.7$\sigma$ \\ \hline
alternative reverberation model & 0.4 & 130 & 2.26 & 1.01 & 1.0 & 2.0 & 0.29 & 1.2$\sigma$ & 7.2$\sigma$ \\
alternative reverberation model & 0.4 & 35 & 2.26 & 1.01 & 1.0 & 2.0 & 0.29 & 8.6$\sigma$ & 8.0$\sigma$
\end{tabular}
\end{table*}

\end{appendix}

\end{document}